\documentclass[%
 reprint,
superscriptaddress,
 amsmath,amssymb,
 aps,
showkeys,
]{revtex4-2}

\usepackage{graphicx}
\usepackage{dcolumn}
\usepackage{bm}
\usepackage{hyperref}
\usepackage{color}
\usepackage{multirow}
\usepackage[table,xcdraw]{xcolor}
\usepackage{tikz}
\usepackage{array}
\newcolumntype{P}[1]{>{\centering\arraybackslash}p{#1}}
\newcolumntype{M}[1]{>{\centering\arraybackslash}m{#1}}

\usepackage{mathtools}
\usepackage{commath}

\usepackage{hyperref}

\begin{document}

\preprint{APS/123-QED}

\title{Selection rules and dynamic magnetoelectric effect of the spin waves in \\multiferroic BiFeO$_3$}

\author{D. G. Farkas}
\affiliation{Department of Physics, Budapest University of Technology and Economics, 1111 Budapest, Hungary}
\affiliation{MTA-BME Condensed Matter Research Group, 1111 Budapest, Hungary}

\author{D. Szaller}
\affiliation{Institute of Solid State Physics, TU Wien, 1040 Vienna, Austria}
\affiliation{Department of Physics, Budapest University of Technology and Economics,  1111 Budapest, Hungary}

\author{I. K{\'e}zsm{\'a}rki}
\affiliation{Department of Physics, Budapest University of Technology and Economics, 1111 Budapest, Hungary}
\affiliation{Experimental Physics V, Center for Electronic Correlations and Magnetism, Institute of Physics, University of Augsburg, 86159 Augsburg, Germany}

\author{U. Nagel}
\affiliation{National Institute of Chemical Physics and Biophysics, Akadeemia tee 23, 12618 Tallinn, Estonia}

\author{T. R{\~o}{\~o}m}
\affiliation{National Institute of Chemical Physics and Biophysics, Akadeemia tee 23, 12618 Tallinn, Estonia}

\author{L. Peedu}
\affiliation{National Institute of Chemical Physics and Biophysics, Akadeemia tee 23, 12618 Tallinn, Estonia}

\author{J. Viirok}
\affiliation{National Institute of Chemical Physics and Biophysics, Akadeemia tee 23, 12618 Tallinn, Estonia}

\author{J. S. White}
\affiliation{Laboratory for Neutron Scattering and Imaging (LNS), Paul Scherrer Institut (PSI), CH-5232 Villigen, Switzerland}

\author{R. Cubitt}
\affiliation{Institut Laue-Langevin, 71 avenue des Martyrs, CS 20156, 38042 Grenoble cedex 9, France}

\author{T. Ito}
\affiliation{National Institute of Advanced Industrial Science and Technology (AIST), Tsukuba, 305-8562 Ibaraki, Japan}

\author{R. S. Fishman}
\affiliation{Materials Science and Technology Division, Oak Ridge National Laboratory, Oak Ridge, Tennessee 37831, USA}

\author{S. Bord{\'a}cs}
\affiliation{Department of Physics, Budapest University of Technology and Economics, 1111 Budapest, Hungary}
\affiliation{Hungarian Academy of Sciences, Premium Postdoctor Program, 1051 Budapest, Hungary}

\date{\today}

\begin{abstract}

We report the magnetic field dependence of the THz absorption and non-reciprocal directional dichroism spectra of BiFeO$_3$ measured on the three principal crystal cuts for fields applied along the three principal directions of each cut. From the systematic study of the light polarization dependence we deduced the optical selection rules of the spin-wave excitations. Our THz data, combined with small-angle neutron scattering results showed that i) an in-plane magnetic field rotates the $\mathbf{q}$ vectors of the cycloids perpendicular to the magnetic field, and ii) the selection rules are mostly determined by the orientation of the $\mathbf{q}$ vector with respect to the electromagnetic fields. We observed a magnetic field history dependent change in the strength and the frequency of the spin-wave modes, which we attributed to the change of the orientation and the length of the cycloidal $\mathbf{q}$ vector, respectively. Finally, we compared our experimental data with the results of linear spin-wave theory that reproduces the magnetic field dependence of the spin-wave frequencies and most of the selection rules, from which we identified the spin-polarization coupling terms relevant for the optical magnetoelectric effect.
\end{abstract}

\keywords{Multiferroics, BiFeO$_3$, spin-wave excitations, THz spectroscopy, non-reciprocal directional dichroism}
\maketitle

\section{\label{sec:intro}Introduction}

Multiferroic materials with coexisting magnetic and ferroelectric orders have received much attention due to their enhanced magnetoelectric coupling mediating interaction between spin and electric polarization \cite{hill2000there, dong2015multiferroic, manfred2016evolution, Spaldin2019}. Among multiferroics, BiFeO$_3$ is the most extensively studied compound as its multiferroic phase is stable above room-temperature \cite{Moreau1971} and it has one of the largest spontaneous ferroelectric polarization even compared to canonical ferroelectrics \cite{Park2014review}, which puts its technical applications within reach. As first steps toward this goal, the operation of spin-valves and ferroelectric tunnel junctions based on BiFeO$_3$ thin films have been demonstrated \cite{Martin2007, Martin2010, Yamada2013, Qu2012, Chakrabarti2014}, and very recently the concept of a magnetoelectric logic device was also proposed \cite{manipatruni2019}.

\begin{figure*}[th!]
\centering
\includegraphics[width=\textwidth]{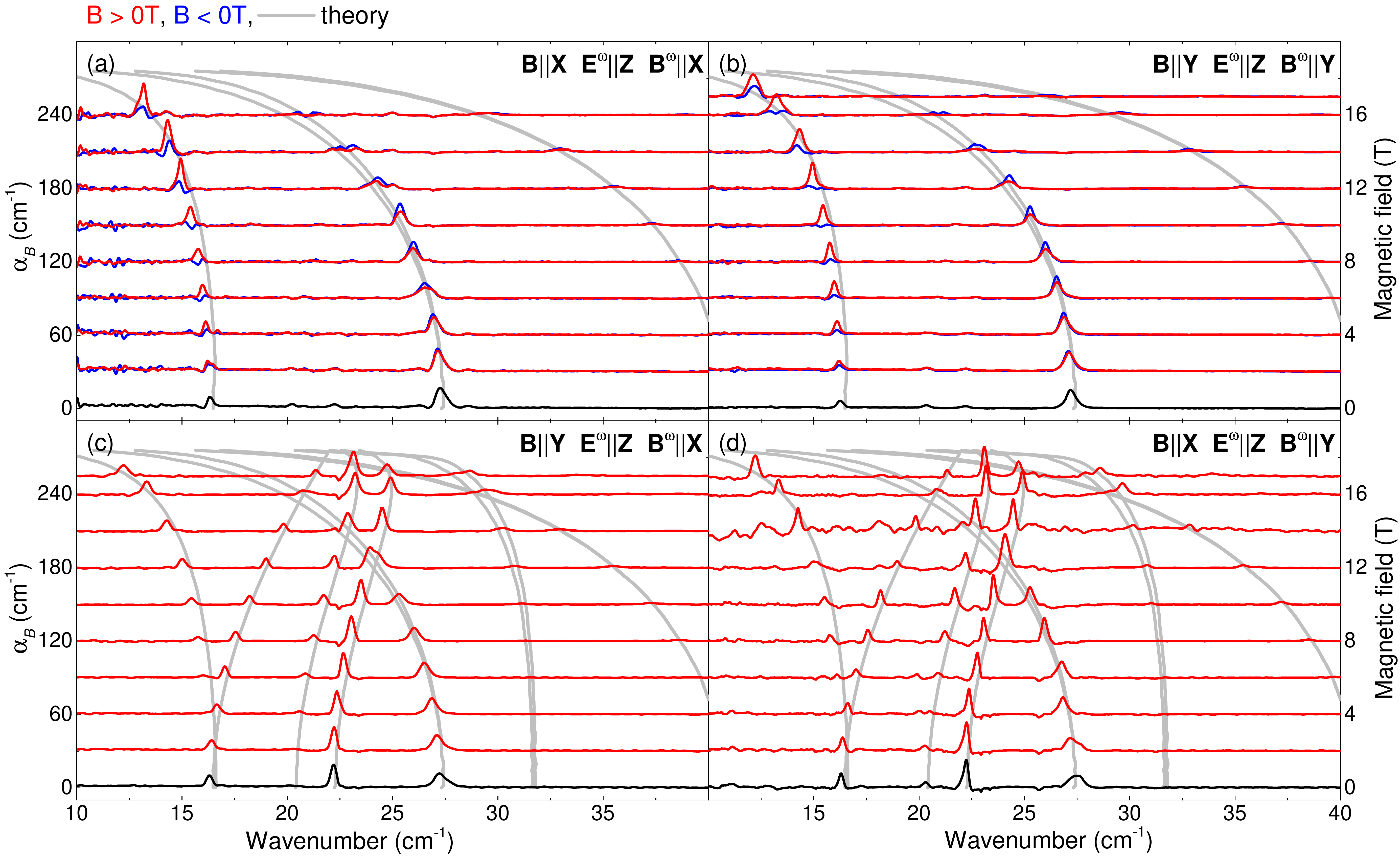}
\caption{Magnetic field dependence of the THz absorption spectra of BiFeO$_3$ at 4\,K. The magnetic field is applied along \textbf{X}, panels (a) and (d), and along \textbf{Y}, panels (b) and (c). THz light polarization is $\mathbf{E^\omega} \parallel \mathbf{Z}$ and $\mathbf{B^\omega} \parallel \mathbf{X}$ in (a) and (c), and $\mathbf{E^\omega} \parallel \mathbf{Z}$ and $\mathbf{B^\omega} \parallel \mathbf{Y}$ in (b) and (d). The spectra are plotted for positive and negative fields in red and blue, respectively, with a vertical offset proportional to the magnitude of the field. Grey lines show the magnetic field dependence of spin-wave frequencies deduced from linear spin-wave theory.}
\label{fig_Bdcplane_Ew111}
\end{figure*}

BiFeO$_3$ has a perovskite crystal structure, where the magnetic Fe$^{3+}$ ion is located in an octahedral cage of O$^{2-}$ anions. Below the ferroelectric phase transition driven by the lone-pair $6s$ electrons of Bi$^{3+}$ at $T_{\text{c}}$=1100\,K \cite{Moreau1971}, the lattice becomes rhombohedrally distorted, lowering the symmetry group to $R3c$. The ferroelectric polarization can appear along any of the eight cubic $\left \langle 111 \right \rangle$ directions \footnote{In this paper the crystallographic directions are described in the pseudo-cubic notation.}, therefore, eight ferroelectric domain states can be formed. Below $T_\mathrm{N}=640$\,K a G-type antiferromagnetic order develops, thus, BiFeO$_3$ becomes multiferroic \cite{Ito2011}. The ferroelectric distortion breaking the inversion symmetry allows a uniform Dzyaloshinskii-Moriya (DM) interaction. This polarization-induced DM interaction creates a cycloidal modulation of the antiferromagnetic order, in which the spins rotate in the plane defined by the ferroelectric polarization $\mathbf{P}$ and the modulation vector, \textbf{q} \cite{Park2014review}. The $\mathbf{q}$ vectors of the cycloids point to one of the three $\left \langle 1 \overline{1}0 \right \rangle$ directions in the plane normal to $\mathbf{P}$, resulting in three magnetic domain states within a ferroelectric domain. The strength of the Heisenberg exchange, the DM interactions and the magnetic anisotropies were determined by a combination of spectroscopic studies such as inelastic neutron \cite{Matsuda2012, Jeong2012} and light scattering experiments \cite{Cazayous2008}, THz absorption spectroscopy \cite{Talbayev2011PRBBiFeO, Nagel2013, Room2020highfield} as well as theoretical modeling \cite{Fishman2013PRB, Fishman2013}.

Until the laser-diode heating floating-zone (LDFZ) technique enabled the growth of large ferroelectric monodomain crystals of BiFeO$_3$, limited information was available on the bulk magnetoelectric effect \cite{Popov1993, Popov2001}. This advance in crystal growth allowed the systematic study of the magnetic field induced excess polarization \cite{Tokunaga2015}. The finite magnetic order induced polarization detected in the (111) plane, perpendicular to the structural polarization is interpreted in terms of an anisotropic inverse DM interaction \cite{Tokunaga2015}. Small-angle neutron scattering (SANS) measurements demonstrated that an external magnetic field exceeding a pinning threshold of $\sim5$\,T reorients the $\mathbf{q}$ vectors to become normal to the magnetic field applied in the (111) plane \cite{Bordacs2018}. This field induced rearrangment of the cycloidal domains was found to be responsible for the non-volatile change in the transverse polarization \cite{Bordacs2018, Tokunaga2015}. A new magnetoelectric term quadratic in spins was suggested to describe the magnetic order induced polarization within the (111) plane \cite{Bordacs2018}, which was confirmed very recently by high-field magneto-current measurements \cite{kawachi2019direct}.

In order to gain more insight into the microscopic interactions responsible for the magnetoelectric coupling in bulk BiFeO$_3$, we performed THz spectroscopy on the three principal cuts of large ferroelectric monodomain single crystals of BiFeO$_3$. Previous THz studies \cite{Talbayev2011PRBBiFeO, Nagel2013} done on $(001)$ face samples provided limited access to the selection rules. Here, by studying the light polarization and magnetic field dependence of the spin-wave excitations we obtained information about the coupling between the oscillating electric and magnetic fields and the cycloidal state. Moreover, in agreement with Ref. \cite{Kezsmarki2015}, we detected strong non-reciprocal directional dichroism (NDD), which is the light absorption difference along and opposite to a given direction. This phenomenon, which arises in this long-wavelength regime only if an excitation is simultaneously electric and magnetic dipole active \cite{Kezsmarki2011, Bordacs2012, takahashi2012, Kezsmarki2015, yu2018, viirok2019directional, Kezsmarki2014}, helped us to determine the relative phase between the oscillating electric and magnetic dipoles. Next, we compared THz spectra and SANS images recorded after the application of magnetic fields with different magnitudes to understand the non-volatile change of the THz absorption and the cycloidal domains. Finally, we analysed the magnetic field dependence of the magnon energies and their intensities in linear spin-wave theory.

\begin{figure*}[th!]
\centering
\includegraphics[width=\textwidth]{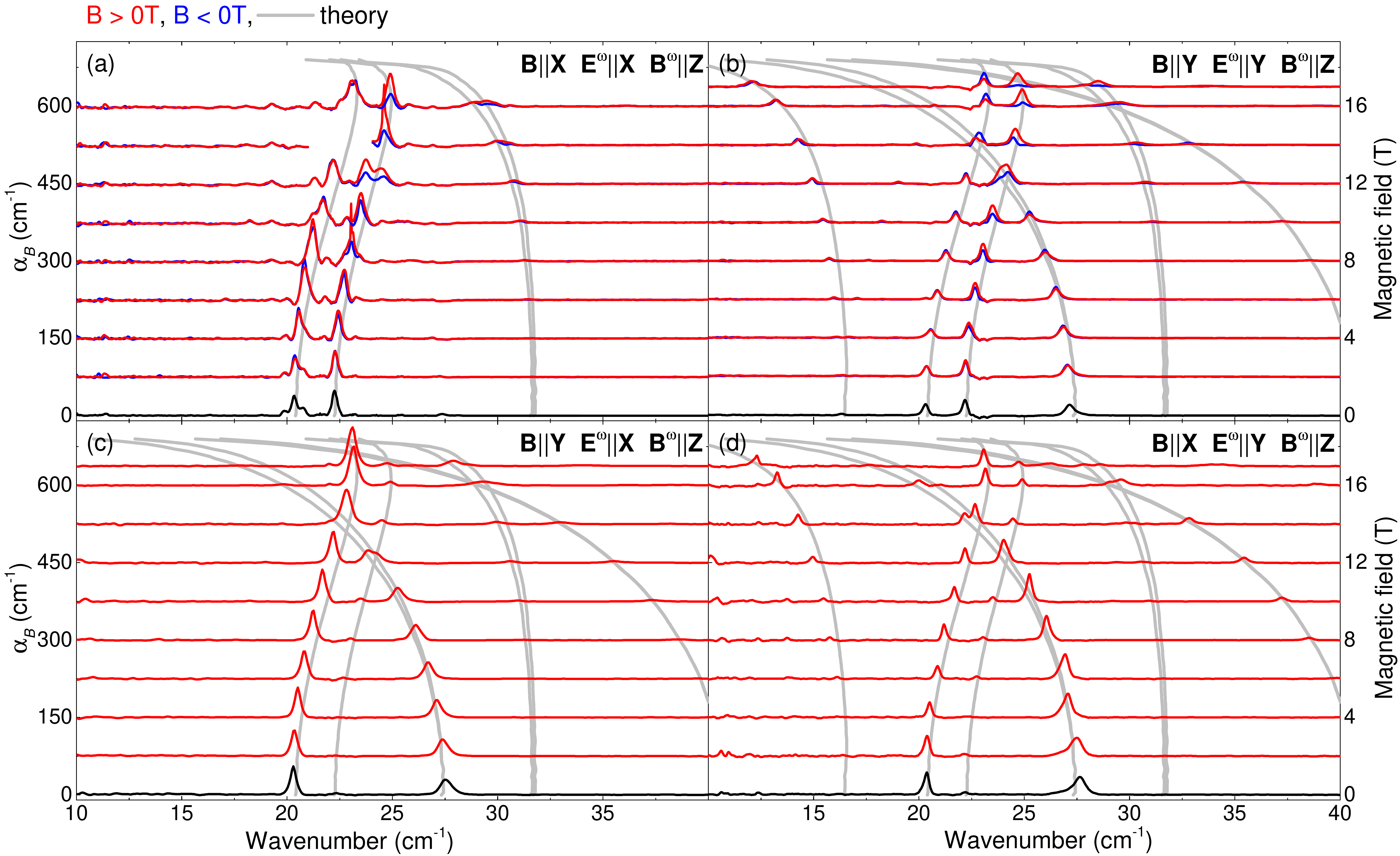}
\caption{Magnetic field dependence of the THz absorption spectra of BiFeO$_3$ at 4\,K. The magnetic field is applied along \textbf{X}, panels (a) and (d), and along \textbf{Y}, panels (b) and (c). THz light polarization is $\mathbf{E^\omega} \parallel \mathbf{X}$ and $\mathbf{B^\omega} \parallel \mathbf{Z}$ in (a) and (c), and $\mathbf{E^\omega} \parallel \mathbf{Y}$ and $\mathbf{B^\omega} \parallel \mathbf{Z}$ in (b) and (d). The spectra are plotted for positive and negative fields in red and blue, respectively, with a vertical offset proportional to the magnitude of the field. Grey lines show the magnetic field dependence of spin-wave frequencies deduced from linear spin-wave theory. Due to strong noise a small frequency window is not displayed in the 14\,T spectrum in panel (a).}
\label{fig_Bdcplane_Bw111}
\end{figure*}

\section{\label{sec:methods}Experimental methods}

A ferroelectric monodomain BiFeO$_3$ single crystal was grown by the LDFZ technique. The details of the crystal growth were reported in Ref.~\onlinecite{Ito2011}. For the optical experiments, samples with large (1$\overline{1}$0), (11$\overline{2}$) and (111) faces were cut to a thickness of about 0.5\,mm with a wedge angle of 2$^{\circ}$ to suppress the interference caused by internal reflections. From now on we will use the following notation for the principal axes throughout this paper: $\mathbf{X} \parallel$ [1$\bar{1}$0], $\mathbf{Y} \parallel$ [11$\bar{2}$] and $\mathbf{Z} \parallel$ [111] where $[hkl]$ are directions in the pseudocubic unit cell reference frame.

Transmission measurements were carried out in the THz range with 0.5\,cm$^{-1}$ resolution using a Martin-Puplett interferometer and a 300\,mK bolometer in NICPB, Tallinn, Estonia. The THz absorption was measured at around 4\,K with normal incidence radiation, polarized linearly along the two high symmetry directions, $\mathbf{X}, \mathbf{Y}$ or $\mathbf{Z}$, lying in the plane of the sample. For each sample the magnetic field was applied either along the propagation direction $\mathbf{k}$, normal to the plane of the sample ($\mathbf{k} \parallel \mathbf{B}$), which is called Faraday configuration, or along the in-plane high symmetry directions, termed as Voigt configuration ($\mathbf{k} \perp \mathbf{B}$). Unless stated otherwise, in order to establish a well defined cycloidal domain configuration, the maximal available field of 17\,T was applied before measuring the magnetic field dependence of the spectra \cite{Bordacs2018}. We note here that this maximal field is below the critical field where the cycloidal order is destroyed \cite{Tokunaga2015}, but large enough to rotate the cycloidal $\mathbf{q}$ vectors \cite{Bordacs2018}.

In the Faraday configuration the sample holder allowed the measurement of the reference intensity spectrum through an empty hole. In this case the $\alpha_s (B)$ absorption spectra were first determined using the following formula: $\alpha_s (B) = - d^{-1} \ln[I_{s} (B) / I_{r} (B)]$, where $d$ is the thickness of the sample, $I_{s} (B)$ and $I_{r} (B)$ are the measured transmission intensities through the sample and the reference hole in an external field $B$, respectively. The two main sources of experimental errors are the small differences in the shape of the sample and the reference hole, which could result in a frequency-dependent baseline due to diffraction  differences, and the residual interference fringes present in the sample spectra despite the sample being wedged. In the Voigt experiments, no reference measurement was possible.

\begin{figure*}[th!]
\centering
\includegraphics[width=\textwidth]{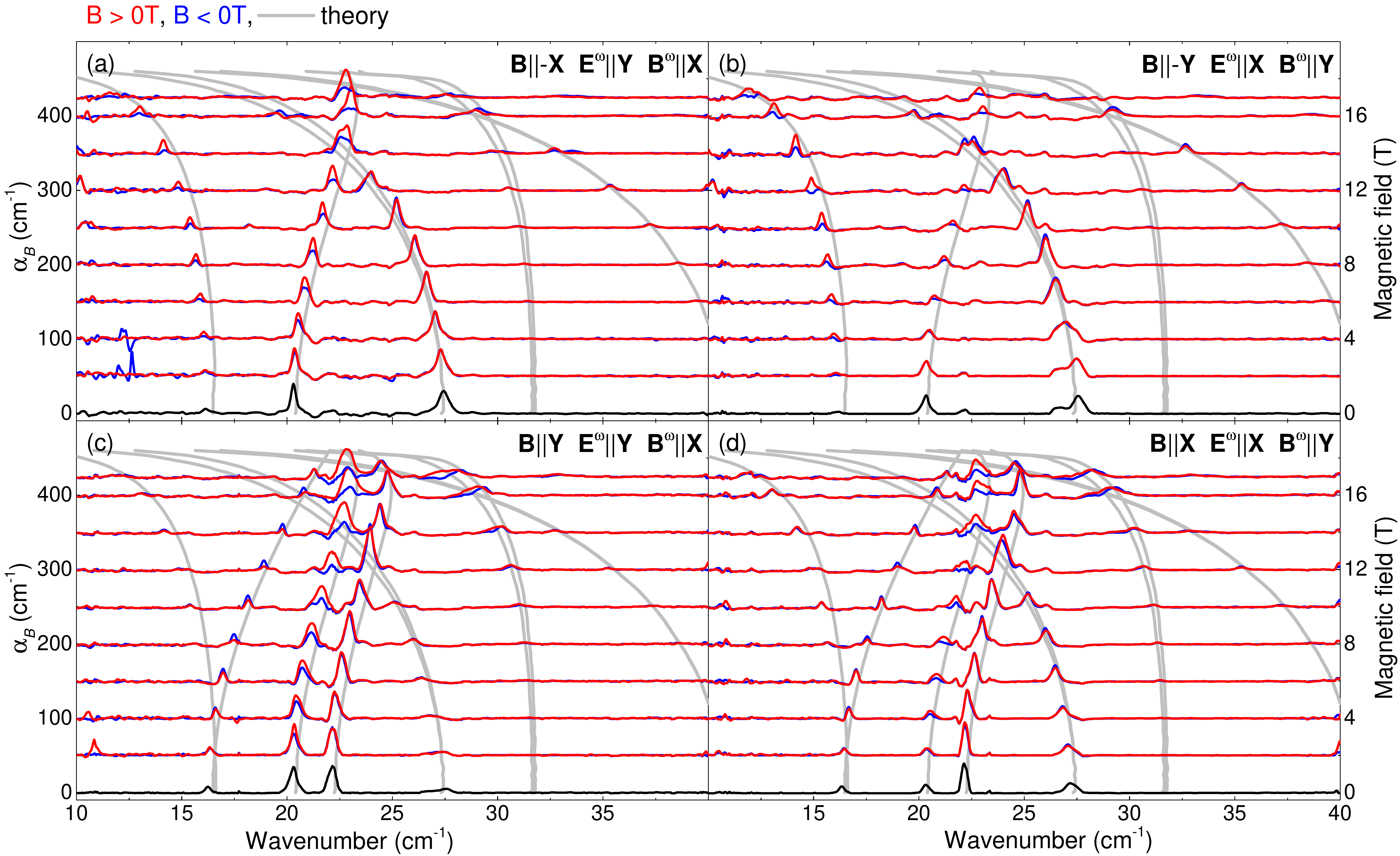}
\caption{Magnetic field dependence of the THz absorption spectra of BiFeO$_3$ at 4\,K. The magnetic field is applied along \textbf{X}, panels (a) and (d), and along \textbf{Y}, panels (b) and (c). THz light polarization is $\mathbf{E^\omega} \parallel \mathbf{Y}$ and $\mathbf{B^\omega} \parallel \mathbf{X}$ in (a) and (c), and $\mathbf{E^\omega} \parallel \mathbf{X}$ and $\mathbf{B^\omega} \parallel \mathbf{Y}$ in (b) and (d). The spectra are plotted for positive and negative fields in red and blue, respectively, with a vertical offset proportional to the magnitude of the field. Grey lines show the magnetic field dependence of spin-wave frequencies deduced from linear spin-wave theory.}
\label{fig_Bdcplane_BwEwinplane}
\end{figure*}

To eliminate these errors, and also to evaluate the results of Voigt measurements we applied the following method to deduce the magnetic field dependent part of the absorption spectrum \cite{Peedu2019LiNiPO, Szaller2017TbFeBO, Szaller2020MnMoO}. First, the field-induced change in the absorption spectrum is calculated as $\Delta \alpha_B (B) = - d^{-1} \ln[I_{s} (B) / I_{s} (0)]$. In this spectrum the magnetic field independent parts are cancelled out and the absorption becomes negative at around the field dependent resonances present already in zero field. The spectra are normalized far from resonance frequencies to compensate for small fluctuations in the intensity of the source. Next, $\alpha_B (0)$ is recovered by calculating the median of the absorption coefficients, $\Delta \alpha_B (B)$, at each frequency and finally $\alpha_B (0)$ is added to $\Delta \alpha_B (B)$ to get the absorption spectrum of features with clear magnetic field dependence: $\alpha_B (B) = \alpha_B (0) + \Delta \alpha_B (B)$.

The above method was used to evaluate all spectra measured either in the Faraday or in the Voigt configuration, and we present the $\alpha_B$ spectra below. This method is applicable in the present case since the spin-wave excitations of BiFeO$_3$ appear as sharp resonances in the absorption spectra and their energies considerably shift with the field. We verified this by comparing the $\alpha_s$ and $\alpha_B$ spectra measured in the Faraday experiments and found no major differences but less fluctuations in the baseline of $\alpha_B$.

SANS experiments were performed using the D33 instrument of the Institut Laue Langevin and the SANS-I instrument of the Paul Scherrer Institute. The incoming neutron wavelength was set to 8 \AA. The zero-field cooled (ZFC) samples were measured in fields up to 8\,T and 10.5\,T in the two institutes, respectively. SANS images were recorded in zero field at various temperatures up to 300\,K. At fixed magnetic field and temperature, a scattering image was obtained by summing the detector images for neutrons incoming nearly along the $\mathrm{Z}$ axis as the sample was rocked (rotated) around the $\mathrm{X}$ and $\mathrm{Y}$ axes between $-3^\circ$ and $+3^\circ$ in $0.2^\circ$ steps. Since the cycloidal structure is stable up to $T_\mathrm{N}=640$\,K there was no possibility to measure a reference image in the paramagnetic phase. To evaluate the magnetic field dependence of the length of the $\mathbf{q}$ vectors the azimuthally integrated SANS intensity was fitted by a Lorentzian function.

\section{\label{sec:resultsinplaneB}THz spectra for magnetic fields applied in the plane normal to Z}

The absorption spectra measured in fields applied in the $\mathrm{XY}$ plane, either along the $\mathrm{X}$ or the $\mathrm{Y}$ direction are presented in Figs.~\ref{fig_Bdcplane_Ew111}-\ref{fig_Bdcplane_BwEwinplane}. In each figure, the panels are arranged into a two-by-two matrix so that the light polarization ($\mathbf{E}^\omega$ and $\mathbf{B}^\omega$) with respect to the crystallographic axes is the same for the columns whereas its orientation relative to the static magnetic field is identical in the rows. The absorption spectra are plotted for positive and negative magnetic fields in red and blue, respectively, with a vertical offset proportional to the magnitude of the field. We define NDD as the difference of the absorption spectra measured in $+B$ and $-B$ fields following Ref.~\onlinecite{Kezsmarki2015}. 

We observed a series of magnetic field dependent modes that we assigned to spin-wave excitations of the cycloidal order in agreement with former Raman and infrared studies \cite{Cazayous2008, Talbayev2011PRBBiFeO, Nagel2013}. Following the notation in Ref.~\onlinecite{Fishman2013PRB}, we label in-plane and out-of-plane excitations as $\Phi_{0, 1 ...}$ and $\Psi_{0, 1 ...}$, respectively. The Goldstone mode, $\Phi_{0}$ is out of  the frequency range of our experiments. The resonance frequencies of the spin-wave modes show the same field dependence for $\mathbf{B} \parallel \mathbf{X}$ and $\mathbf{B} \parallel \mathbf{Y}$ as demonstrated in Figs.~\ref{fig_Bdcplane_Ew111}-\ref{fig_Bdcplane_BwEwinplane}. This result implies that the static fields applied in the $\mathrm{XY}$ plane  distort the cycloids in the same way irrespective of the direction of their $\mathbf{q}$ vector.

\begin{table*}[th!]
\setlength{\tabcolsep}{0pt}
\centering
\begin{tabular}{|P{1.5cm}|P{0.9cm}||P{0.7cm}|P{0.7cm}|P{0.7cm}|P{0.7cm}||P{0.7cm}|P{0.7cm}|P{0.7cm}|P{0.7cm}||P{0.7cm}|P{0.7cm}|P{0.7cm}|P{0.7cm}|}
\hline
\multicolumn{2}{|c||}{\textbf{Mode notation}} & \multicolumn{4}{c||}{\textbf{B$^\omega$} $\parallel$ \textbf{Z}} & \multicolumn{4}{c||}{\textbf{B$^\omega$} $\parallel$ \textbf{q}} & \multicolumn{4}{c|}{\textbf{B$^\omega$} $\parallel$ \textbf{Z x q}} \\
\hline
\multicolumn{2}{|c||}{\textbf{and zero field}} & \multicolumn{2}{c|}{\textbf{Exp.}} & \multicolumn{2}{c||}{\textbf{Model}} & \multicolumn{2}{c|}{\textbf{Exp.}} & \multicolumn{2}{c||}{\textbf{Model}} & \multicolumn{2}{c|}{\textbf{Exp.}} & \multicolumn{2}{c|}{\textbf{Model}} \\
\hline
\multicolumn{2}{|c||}{\textbf{frequency}} & B=0 & B$\neq$0 & B=0 & B$\neq$0 & B=0 & B$\neq$0 & B=0 & B$\neq$0 & B=0 & B$\neq$0 & B=0 & B$\neq$0 \\
\hline\hline
16.3 cm$^{-1}$ & $\Phi_{1}^{(1)}$ & \cellcolor[HTML]{BF8F00}X & \cellcolor[HTML]{BF8F00}X & \cellcolor[HTML]{BF8F00}X & \cellcolor[HTML]{BF8F00}X & \cellcolor[HTML]{BF8F00}X & & \cellcolor[HTML]{BF8F00}X & \cellcolor[HTML]{BF8F00}X & \cellcolor[HTML]{00B050}\checkmark & \cellcolor[HTML]{00B050}\checkmark & \cellcolor[HTML]{00B050}\checkmark & \cellcolor[HTML]{00B050}\checkmark \\
\hline
16.3 cm$^{-1}$ & $\Psi_{0}$ & \cellcolor[HTML]{BF8F00}X & \cellcolor[HTML]{BF8F00}X & \cellcolor[HTML]{BF8F00}X & \cellcolor[HTML]{BF8F00}X & \cellcolor[HTML]{00B050}\checkmark & \cellcolor[HTML]{00B050}\checkmark & \cellcolor[HTML]{00B050}\checkmark & \cellcolor[HTML]{00B050}\checkmark & \cellcolor[HTML]{BF8F00}X & \cellcolor[HTML]{BF8F00}X & \cellcolor[HTML]{BF8F00}X & \cellcolor[HTML]{BF8F00}X \\
\hline
20.3 cm$^{-1}$ & $\Psi_{1}^{(2)}$ & & & \cellcolor[HTML]{00B050}\checkmark & \cellcolor[HTML]{00B050}\checkmark & \cellcolor[HTML]{BF8F00}X & \cellcolor[HTML]{00B050}\checkmark & \cellcolor[HTML]{BF8F00}X & \cellcolor[HTML]{00B050}\checkmark & \cellcolor[HTML]{BF8F00}X & \cellcolor[HTML]{BF8F00}X & \cellcolor[HTML]{BF8F00}X & \cellcolor[HTML]{BF8F00}X \\
\hline
22.2 cm$^{-1}$ & $\Psi_{1}^{(1)}$ & \cellcolor[HTML]{BF8F00}X & \cellcolor[HTML]{00B050}\checkmark & \cellcolor[HTML]{BF8F00}X & \cellcolor[HTML]{00B050}\checkmark & \cellcolor[HTML]{00B050}\checkmark & \cellcolor[HTML]{00B050}\checkmark & \cellcolor[HTML]{00B050}\checkmark & \cellcolor[HTML]{00B050}\checkmark & \cellcolor[HTML]{BF8F00}X & \cellcolor[HTML]{BF8F00}X & \cellcolor[HTML]{BF8F00}X & \cellcolor[HTML]{BF8F00}X \\
\hline
27.3 cm$^{-1}$ & $\Phi_{2}^{(1,2)}$ & & & \cellcolor[HTML]{BF8F00}X & \cellcolor[HTML]{BF8F00}X & & & \cellcolor[HTML]{BF8F00}X & \cellcolor[HTML]{BF8F00}X & & \cellcolor[HTML]{00B050}\checkmark & \cellcolor[HTML]{00B050}\checkmark & \cellcolor[HTML]{00B050}\checkmark \\
\hline
31.7 cm$^{-1}$ & $\Psi_{2}^{(1,2)}$ & \cellcolor[HTML]{BF8F00}X & & \cellcolor[HTML]{BF8F00}X & \cellcolor[HTML]{00B050}\checkmark & \cellcolor[HTML]{BF8F00}X & \cellcolor[HTML]{00B050}\checkmark & \cellcolor[HTML]{BF8F00}X & \cellcolor[HTML]{00B050}\checkmark & \cellcolor[HTML]{BF8F00}X & \cellcolor[HTML]{BF8F00}X & \cellcolor[HTML]{BF8F00}X & \cellcolor[HTML]{BF8F00}X \\
\hline
40.5 cm$^{-1}$ & $\Phi_{3}^{(1,2)}$ & \cellcolor[HTML]{BF8F00}X & & \cellcolor[HTML]{BF8F00}X & \cellcolor[HTML]{BF8F00}X & \cellcolor[HTML]{BF8F00}X & & \cellcolor[HTML]{BF8F00}X & \cellcolor[HTML]{BF8F00}X & \cellcolor[HTML]{BF8F00}X & & \cellcolor[HTML]{BF8F00}X & \cellcolor[HTML]{00B050}\checkmark \\
\hline
\end{tabular}
\caption{Magnetic dipole selection rules of the spin-wave excitations derived in zero ($B=0$) and finite ($B\neq0$) magnetic fields. $\checkmark$/X on green/brown background indicates an active/silent mode. The directions of the oscillating magnetic field are given with respect to the cycloidal order described by the wave vector $\mathbf{q}$. The experimental results (Exp.) are compared to the calculated magnetization matrix elements (Model).}
\label{table:selectionrules_B}
\end{table*}

\begin{table*}[th!]
\setlength{\tabcolsep}{0pt}
\centering
\begin{tabular}{|P{1.5cm}|P{0.9cm}||P{0.7cm}|P{0.7cm}||P{0.7cm}|P{0.7cm}|P{0.7cm}|P{0.7cm}|P{0.7cm}|P{0.7cm}|P{0.7cm}|P{0.7cm}|}
\hline
\multicolumn{2}{|c||}{\textbf{Mode notation}} & \multicolumn{10}{c|}{\textbf{E$^\omega$} $\parallel$ \textbf{Z}} \\
\hline
\multicolumn{2}{|c||}{\textbf{and zero field}} & \multicolumn{2}{c||}{\textbf{Exp.}} & \multicolumn{2}{c|}{\textbf{SC(1)$_Y$}} & \multicolumn{2}{c|}{\textbf{SC(2)$_Z$}} & \multicolumn{2}{c|}{\textbf{MS(1)$_Z$}} & \multicolumn{2}{c|}{\textbf{ANI(3)$_Z$}} \\
\hline
\multicolumn{2}{|c||}{\textbf{frequency}} & B=0 & B$\neq$0 & B=0 & B$\neq$0 & B=0 & B$\neq$0 & B=0 & B$\neq$0 & B=0 & B$\neq$0\\
\hline\hline
16.3 cm$^{-1}$ & $\Phi_{1}^{(1)}$ & \cellcolor[HTML]{BF8F00}X & \cellcolor[HTML]{00B050}\checkmark & \cellcolor[HTML]{BF8F00}X & \cellcolor[HTML]{BF8F00}X & \cellcolor[HTML]{BF8F00}X & \cellcolor[HTML]{00B050}\checkmark & \cellcolor[HTML]{BF8F00}X & \cellcolor[HTML]{00B050}\checkmark & \cellcolor[HTML]{BF8F00}X & \cellcolor[HTML]{00B050}\checkmark \\
\hline
16.3 cm$^{-1}$ & $\Psi_{0}$ & \cellcolor[HTML]{BF8F00}X & \cellcolor[HTML]{BF8F00}X & \cellcolor[HTML]{BF8F00}X & \cellcolor[HTML]{BF8F00}X & \cellcolor[HTML]{BF8F00}X & \cellcolor[HTML]{BF8F00}X & \cellcolor[HTML]{BF8F00}X & \cellcolor[HTML]{BF8F00}X & \cellcolor[HTML]{BF8F00}X & \cellcolor[HTML]{BF8F00}X \\
\hline
20.3 cm$^{-1}$ & $\Psi_{1}^{(2)}$ & \cellcolor[HTML]{BF8F00}X & \cellcolor[HTML]{BF8F00}X & \cellcolor[HTML]{BF8F00}X & \cellcolor[HTML]{BF8F00}X & \cellcolor[HTML]{BF8F00}X & \cellcolor[HTML]{BF8F00}X & \cellcolor[HTML]{BF8F00}X & \cellcolor[HTML]{BF8F00}X & \cellcolor[HTML]{BF8F00}X & \cellcolor[HTML]{BF8F00}X \\
\hline
22.2 cm$^{-1}$ & $\Psi_{1}^{(1)}$ & \cellcolor[HTML]{BF8F00}X & \cellcolor[HTML]{BF8F00}X & \cellcolor[HTML]{BF8F00}X & \cellcolor[HTML]{BF8F00}X & \cellcolor[HTML]{BF8F00}X & \cellcolor[HTML]{BF8F00}X & \cellcolor[HTML]{BF8F00}X & \cellcolor[HTML]{BF8F00}X & \cellcolor[HTML]{BF8F00}X & \cellcolor[HTML]{BF8F00}X \\
\hline
27.3 cm$^{-1}$ & $\Phi_{2}^{(1,2)}$ & & \cellcolor[HTML]{00B050}\checkmark & \cellcolor[HTML]{BF8F00}X & \cellcolor[HTML]{BF8F00}X & \cellcolor[HTML]{00B050}\checkmark & \cellcolor[HTML]{00B050}\checkmark & \cellcolor[HTML]{00B050}\checkmark & \cellcolor[HTML]{00B050}\checkmark & \cellcolor[HTML]{00B050}\checkmark & \cellcolor[HTML]{00B050}\checkmark \\
\hline
31.7 cm$^{-1}$ & $\Psi_{2}^{(1,2)}$ & \cellcolor[HTML]{BF8F00}X & \cellcolor[HTML]{BF8F00}X & \cellcolor[HTML]{BF8F00}X & \cellcolor[HTML]{BF8F00}X & \cellcolor[HTML]{BF8F00}X & \cellcolor[HTML]{BF8F00}X & \cellcolor[HTML]{BF8F00}X & \cellcolor[HTML]{BF8F00}X & \cellcolor[HTML]{BF8F00}X & \cellcolor[HTML]{BF8F00}X \\
\hline
40.5 cm$^{-1}$ & $\Phi_{3}^{(1,2)}$ & \cellcolor[HTML]{BF8F00}X & & \cellcolor[HTML]{BF8F00}X & \cellcolor[HTML]{BF8F00}X & \cellcolor[HTML]{BF8F00}X & \cellcolor[HTML]{BF8F00}X & \cellcolor[HTML]{BF8F00}X & \cellcolor[HTML]{BF8F00}X & \cellcolor[HTML]{BF8F00}X & \cellcolor[HTML]{00B050}\checkmark \\
\hline
\end{tabular}
\caption{Electric dipole selection rules of the spin-wave excitations for $\mathbf{E^\omega} \parallel \mathbf{Z}$ derived in zero ($B=0$) and finite ($B\neq0$) magnetic fields. $\checkmark$/X on green/brown background indicates an active/silent  mode. The experimental results (Exp.) are compared to the polarization matrix elements derived from the different spin-polarization coupling terms (for details see the text).}
\label{table:selectionrules_EZ}
\end{table*}

\begin{table*}[th!]
\setlength{\tabcolsep}{0pt}
\centering
\begin{tabular}{|P{1.5cm}|P{0.9cm}||P{0.7cm}|P{0.7cm}||P{0.7cm}|P{0.7cm}|P{0.7cm}|P{0.7cm}|P{0.7cm}|P{0.7cm}|P{0.7cm}|P{0.7cm}|P{0.7cm}|P{0.7cm}|P{0.7cm}|P{0.7cm}|}
\hline
\multicolumn{2}{|c||}{\textbf{Mode notation}} & \multicolumn{2}{c||}{\textbf{E$^\omega$} $\parallel$ \textbf{q}} & \multicolumn{12}{c|}{\textbf{E$^\omega$} $\parallel$ \textbf{X} which is $\parallel$ \textbf{q}} \\
\hline
\multicolumn{2}{|c||}{\textbf{and zero field}} & \multicolumn{2}{c||}{\textbf{Exp.}} & \multicolumn{2}{c|}{\textbf{SC(1)$_X$}} & \multicolumn{2}{c|}{\textbf{SC(2)$_X$}} & \multicolumn{2}{c|}{\textbf{ANI(1)$_X$}} & \multicolumn{2}{c|}{\textbf{ANI(2)$_X$}} & \multicolumn{2}{c|}{\textbf{ANI(4)$_X$}} & \multicolumn{2}{c|}{\textbf{ANI(5)$_X$}}\\
\hline
\multicolumn{2}{|c||}{\textbf{frequency}} & B=0 & B$\neq$0 & B=0 & B$\neq$0 & B=0 & B$\neq$0 & B=0 & B$\neq$0 & B=0 & B$\neq$0 & B=0 & B$\neq$0 & B=0 & B$\neq$0 \\
\hline\hline
16.3 cm$^{-1}$ & $\Phi_{1}^{(1)}$ & \cellcolor[HTML]{BF8F00}X & & \cellcolor[HTML]{BF8F00}X & \cellcolor[HTML]{BF8F00}X & \cellcolor[HTML]{00B050}\checkmark & \cellcolor[HTML]{00B050}\checkmark & \cellcolor[HTML]{BF8F00}X & \cellcolor[HTML]{BF8F00}X & \cellcolor[HTML]{BF8F00}X & \cellcolor[HTML]{BF8F00}X & \cellcolor[HTML]{00B050}\checkmark & \cellcolor[HTML]{00B050}\checkmark & \cellcolor[HTML]{BF8F00}X & \cellcolor[HTML]{BF8F00}X \\
\hline
16.3 cm$^{-1}$ & $\Psi_{0}$ & \cellcolor[HTML]{BF8F00}X & \cellcolor[HTML]{BF8F00}X & \cellcolor[HTML]{BF8F00}X & \cellcolor[HTML]{00B050}\checkmark & \cellcolor[HTML]{BF8F00}X & \cellcolor[HTML]{BF8F00}X & \cellcolor[HTML]{BF8F00}X & \cellcolor[HTML]{BF8F00}X & \cellcolor[HTML]{BF8F00}X & \cellcolor[HTML]{00B050}\checkmark & \cellcolor[HTML]{BF8F00}X & \cellcolor[HTML]{BF8F00}X & \cellcolor[HTML]{BF8F00}X & \cellcolor[HTML]{00B050}\checkmark \\
\hline
20.3 cm$^{-1}$ & $\Psi_{1}^{(2)}$ & \cellcolor[HTML]{00B050}\checkmark$_x$ & \cellcolor[HTML]{00B050}\checkmark$_x$ & \cellcolor[HTML]{00B050}\checkmark & \cellcolor[HTML]{00B050}\checkmark & \cellcolor[HTML]{BF8F00}X & \cellcolor[HTML]{BF8F00}X & \cellcolor[HTML]{BF8F00}X & \cellcolor[HTML]{BF8F00}X & \cellcolor[HTML]{00B050}\checkmark & \cellcolor[HTML]{00B050}\checkmark & \cellcolor[HTML]{BF8F00}X & \cellcolor[HTML]{BF8F00}X & \cellcolor[HTML]{00B050}\checkmark & \cellcolor[HTML]{00B050}\checkmark \\
\hline
22.2 cm$^{-1}$ & $\Psi_{1}^{(1)}$ & \cellcolor[HTML]{BF8F00}X & \cellcolor[HTML]{BF8F00}X & \cellcolor[HTML]{BF8F00}X & \cellcolor[HTML]{BF8F00}X & \cellcolor[HTML]{BF8F00}X & \cellcolor[HTML]{BF8F00}X & \cellcolor[HTML]{BF8F00}X & \cellcolor[HTML]{BF8F00}X & \cellcolor[HTML]{BF8F00}X & \cellcolor[HTML]{BF8F00}X & \cellcolor[HTML]{BF8F00}X & \cellcolor[HTML]{BF8F00}X & \cellcolor[HTML]{BF8F00}X & \cellcolor[HTML]{00B050}\checkmark \\
\hline
27.3 cm$^{-1}$ & $\Phi_{2}^{(1,2)}$ & \cellcolor[HTML]{00B050}\checkmark & \cellcolor[HTML]{00B050}\checkmark & \cellcolor[HTML]{BF8F00}X & \cellcolor[HTML]{BF8F00}X & \cellcolor[HTML]{00B050}\checkmark & \cellcolor[HTML]{00B050}\checkmark & \cellcolor[HTML]{00B050}\checkmark & \cellcolor[HTML]{00B050}\checkmark & \cellcolor[HTML]{BF8F00}X & \cellcolor[HTML]{BF8F00}X & \cellcolor[HTML]{00B050}\checkmark & \cellcolor[HTML]{00B050}\checkmark & \cellcolor[HTML]{BF8F00}X & \cellcolor[HTML]{BF8F00}X \\
\hline
31.7 cm$^{-1}$ & $\Psi_{2}^{(1,2)}$ & \cellcolor[HTML]{BF8F00}X & \cellcolor[HTML]{00B050}\checkmark & \cellcolor[HTML]{BF8F00}X & \cellcolor[HTML]{00B050}\checkmark & \cellcolor[HTML]{BF8F00}X & \cellcolor[HTML]{BF8F00}X & \cellcolor[HTML]{BF8F00}X & \cellcolor[HTML]{BF8F00}X & \cellcolor[HTML]{BF8F00}X & \cellcolor[HTML]{00B050}\checkmark & \cellcolor[HTML]{BF8F00}X & \cellcolor[HTML]{BF8F00}X & \cellcolor[HTML]{00B050}\checkmark & \cellcolor[HTML]{00B050}\checkmark \\
\hline
40.5 cm$^{-1}$ & $\Phi_{3}^{(1,2)}$ & \cellcolor[HTML]{BF8F00}X & \cellcolor[HTML]{00B050}\checkmark & \cellcolor[HTML]{BF8F00}X & \cellcolor[HTML]{BF8F00}X & \cellcolor[HTML]{BF8F00}X & \cellcolor[HTML]{00B050}\checkmark & \cellcolor[HTML]{BF8F00}X & \cellcolor[HTML]{00B050}\checkmark & \cellcolor[HTML]{BF8F00}X & \cellcolor[HTML]{BF8F00}X & \cellcolor[HTML]{BF8F00}X & \cellcolor[HTML]{00B050}\checkmark & \cellcolor[HTML]{BF8F00}X & \cellcolor[HTML]{BF8F00}X \\
\hline
\end{tabular}
\caption{Electric dipole selection rules of the spin-wave excitations for $\mathbf{E^\omega} \parallel \mathbf{q}$ derived in zero ($B=0$) and finite ($B\neq0$) magnetic fields. $\checkmark$/X on green/brown background indicates an active/silent mode. For some resonances, the orientation of the oscillating electric field with respect to a crystallographic axis determine if it is active. In this case, $\checkmark_x$ labels resonances active only for $\mathbf{E^\omega} \parallel \mathbf{X}$. The experimental results (Exp.) are compared to the polarization matrix elements derived from the different spin-polarization coupling terms (for details see the text). In the model calculations, we choose $\mathbf{X} \parallel \mathbf{q}$, thus, we include the experimental selection rules for both $\mathbf{E}^\omega$ $\parallel$ $\mathbf{X}$ and $\mathbf{E}^\omega$ $\parallel$ $\mathbf{q}$ in the same table as the theoretical results.}
\label{table:selectionrules_Eq}
\end{table*}

\begin{table*}[th!]
\setlength{\tabcolsep}{0pt}
\centering
\begin{tabular}{|P{1.5cm}|P{0.9cm}||P{0.85cm}|P{0.85cm}||P{0.7cm}|P{0.7cm}|P{0.7cm}|P{0.7cm}|P{0.7cm}|P{0.7cm}|P{0.7cm}|P{0.7cm}|P{0.7cm}|P{0.7cm}|P{0.7cm}|P{0.7cm}|P{0.7cm}|P{0.7cm}|P{0.7cm}|P{0.7cm}|P{0.7cm}|P{0.7cm}|}
\hline
\multicolumn{2}{|c||}{\textbf{Mode notation}} & \multicolumn{2}{c||}{\textbf{E$^\omega$} $\parallel$ \textbf{\textbf{Z x q}}} & \multicolumn{18}{c|}{\textbf{E$^\omega$} $\parallel$ \textbf{Y} which is $\parallel$ \textbf{Z x q}} \\
\hline
\multicolumn{2}{|c||}{\textbf{and zero field}} & \multicolumn{2}{c||}{\textbf{Exp.}} & \multicolumn{2}{c|}{\textbf{SC(1)$_Y$}} & \multicolumn{2}{c|}{\textbf{SC(1)$_Z$}} & \multicolumn{2}{c|}{\textbf{SC(2)$_Y$}} & \multicolumn{2}{c|}{\textbf{MS(1)$_Y$}} & \multicolumn{2}{c|}{\textbf{MS(2)$_Y$}} & \multicolumn{2}{c|}{\textbf{ANI(1)$_Y$}} & \multicolumn{2}{c|}{\textbf{ANI(2)$_Y$}} & \multicolumn{2}{c|}{\textbf{ANI(4)$_Y$}} & \multicolumn{2}{c|}{\textbf{ANI(5)$_Y$}} \\
\hline
\multicolumn{2}{|c||}{\textbf{frequency}} & B=0 & B$\neq$0 & B=0 & B$\neq$0 & B=0 & B$\neq$0 & B=0 & B$\neq$0 & B=0 & B$\neq$0 & B=0 & B$\neq$0 & B=0 & B$\neq$0 & B=0 & B$\neq$0 & B=0 & B$\neq$0 & B=0 & B$\neq$0\\
\hline\hline
16.3 cm$^{-1}$ & $\Phi_{1}^{(1)}$ & \cellcolor[HTML]{BF8F00}X & \cellcolor[HTML]{00B050}\checkmark$_y$ & \cellcolor[HTML]{BF8F00}X & \cellcolor[HTML]{BF8F00}X & \cellcolor[HTML]{BF8F00}X & \cellcolor[HTML]{BF8F00}X & \cellcolor[HTML]{BF8F00}X & \cellcolor[HTML]{BF8F00}X & \cellcolor[HTML]{BF8F00}X & \cellcolor[HTML]{00B050}\checkmark & \cellcolor[HTML]{BF8F00}X & \cellcolor[HTML]{BF8F00}X & \cellcolor[HTML]{BF8F00}X & \cellcolor[HTML]{BF8F00}X & \cellcolor[HTML]{BF8F00}X & \cellcolor[HTML]{00B050}\checkmark & \cellcolor[HTML]{BF8F00}X & \cellcolor[HTML]{BF8F00}X & \cellcolor[HTML]{BF8F00}X & \cellcolor[HTML]{00B050}\checkmark \\
\hline
16.3 cm$^{-1}$ & $\Psi_{0}$ & \cellcolor[HTML]{BF8F00}X & \cellcolor[HTML]{BF8F00}X & \cellcolor[HTML]{BF8F00}X & \cellcolor[HTML]{BF8F00}X & \cellcolor[HTML]{BF8F00}X & \cellcolor[HTML]{BF8F00}X & \cellcolor[HTML]{00B050}\checkmark & \cellcolor[HTML]{00B050}\checkmark & \cellcolor[HTML]{BF8F00}X & \cellcolor[HTML]{BF8F00}X & \cellcolor[HTML]{00B050}\checkmark & \cellcolor[HTML]{00B050}\checkmark & \cellcolor[HTML]{BF8F00}X & \cellcolor[HTML]{BF8F00}X & \cellcolor[HTML]{BF8F00}X & \cellcolor[HTML]{BF8F00}X & \cellcolor[HTML]{00B050}\checkmark & \cellcolor[HTML]{00B050}\checkmark & \cellcolor[HTML]{BF8F00}X & \cellcolor[HTML]{BF8F00}X \\
\hline
20.3 cm$^{-1}$ & $\Psi_{1}^{(2)}$ & \cellcolor[HTML]{00B050}\checkmark$_y$ & \cellcolor[HTML]{00B050}\checkmark$_y$ & \cellcolor[HTML]{BF8F00}X & \cellcolor[HTML]{BF8F00}X & \cellcolor[HTML]{BF8F00}X & \cellcolor[HTML]{BF8F00}X & \cellcolor[HTML]{BF8F00}X & \cellcolor[HTML]{00B050}\checkmark & \cellcolor[HTML]{BF8F00}X & \cellcolor[HTML]{BF8F00}X & \cellcolor[HTML]{BF8F00}X & \cellcolor[HTML]{00B050}\checkmark & \cellcolor[HTML]{BF8F00}X & \cellcolor[HTML]{BF8F00}X & \cellcolor[HTML]{BF8F00}X & \cellcolor[HTML]{BF8F00}X & \cellcolor[HTML]{BF8F00}X & \cellcolor[HTML]{00B050}\checkmark & \cellcolor[HTML]{BF8F00}X & \cellcolor[HTML]{BF8F00}X \\
\hline
22.2 cm$^{-1}$ & $\Psi_{1}^{(1)}$ & \cellcolor[HTML]{00B050}\checkmark & \cellcolor[HTML]{00B050}\checkmark & \cellcolor[HTML]{BF8F00}X & \cellcolor[HTML]{BF8F00}X & \cellcolor[HTML]{00B050}\checkmark & \cellcolor[HTML]{00B050}\checkmark & \cellcolor[HTML]{00B050}\checkmark & \cellcolor[HTML]{00B050}\checkmark & \cellcolor[HTML]{BF8F00}X & \cellcolor[HTML]{BF8F00}X & \cellcolor[HTML]{00B050}\checkmark & \cellcolor[HTML]{00B050}\checkmark & \cellcolor[HTML]{00B050}\checkmark & \cellcolor[HTML]{00B050}\checkmark & \cellcolor[HTML]{BF8F00}X & \cellcolor[HTML]{BF8F00}X & \cellcolor[HTML]{00B050}\checkmark & \cellcolor[HTML]{00B050}\checkmark & \cellcolor[HTML]{BF8F00}X & \cellcolor[HTML]{BF8F00}X \\
\hline
27.3 cm$^{-1}$ & $\Phi_{2}^{(1,2)}$ & \cellcolor[HTML]{00B050}\checkmark$_y$ & \cellcolor[HTML]{00B050}\checkmark$_y$ & \cellcolor[HTML]{BF8F00}X & \cellcolor[HTML]{BF8F00}X & \cellcolor[HTML]{BF8F00}X & \cellcolor[HTML]{BF8F00}X & \cellcolor[HTML]{BF8F00}X & \cellcolor[HTML]{BF8F00}X & \cellcolor[HTML]{00B050}\checkmark & \cellcolor[HTML]{00B050}\checkmark & \cellcolor[HTML]{BF8F00}X & \cellcolor[HTML]{BF8F00}X & \cellcolor[HTML]{BF8F00}X & \cellcolor[HTML]{BF8F00}X & \cellcolor[HTML]{00B050}\checkmark & \cellcolor[HTML]{00B050}\checkmark & \cellcolor[HTML]{BF8F00}X & \cellcolor[HTML]{BF8F00}X & \cellcolor[HTML]{00B050}\checkmark & \cellcolor[HTML]{00B050}\checkmark \\
\hline
31.7 cm$^{-1}$ & $\Psi_{2}^{(1,2)}$ & \cellcolor[HTML]{BF8F00}X & & \cellcolor[HTML]{BF8F00}X & \cellcolor[HTML]{BF8F00}X & \cellcolor[HTML]{BF8F00}X & \cellcolor[HTML]{00B050}\checkmark & \cellcolor[HTML]{00B050}\checkmark & \cellcolor[HTML]{00B050}\checkmark & \cellcolor[HTML]{BF8F00}X & \cellcolor[HTML]{BF8F00}X & \cellcolor[HTML]{00B050}\checkmark & \cellcolor[HTML]{00B050}\checkmark & \cellcolor[HTML]{BF8F00}X & \cellcolor[HTML]{00B050}\checkmark & \cellcolor[HTML]{BF8F00}X & \cellcolor[HTML]{BF8F00}X & \cellcolor[HTML]{00B050}\checkmark & \cellcolor[HTML]{00B050}\checkmark & \cellcolor[HTML]{BF8F00}X & \cellcolor[HTML]{BF8F00}X \\
\hline
40.5 cm$^{-1}$ & $\Phi_{3}^{(1,2)}$ & \cellcolor[HTML]{BF8F00}X & \cellcolor[HTML]{00B050}\checkmark$_y$ & \cellcolor[HTML]{BF8F00}X & \cellcolor[HTML]{BF8F00}X & \cellcolor[HTML]{BF8F00}X & \cellcolor[HTML]{BF8F00}X & \cellcolor[HTML]{BF8F00}X & \cellcolor[HTML]{BF8F00}X & \cellcolor[HTML]{BF8F00}X & \cellcolor[HTML]{BF8F00}X & \cellcolor[HTML]{BF8F00}X & \cellcolor[HTML]{BF8F00}X & \cellcolor[HTML]{BF8F00}X & \cellcolor[HTML]{BF8F00}X & \cellcolor[HTML]{BF8F00}X & \cellcolor[HTML]{00B050}\checkmark & \cellcolor[HTML]{BF8F00}X & \cellcolor[HTML]{BF8F00}X & \cellcolor[HTML]{BF8F00}X & \cellcolor[HTML]{00B050}\checkmark \\
\hline
\end{tabular}
\caption{Electric dipole selection rules of the spin-wave excitations for $\mathbf{E^\omega} \parallel \mathbf{Z} \times \mathbf{q}$ derived in zero ($B=0$) and finite ($B\neq0$) magnetic fields. $\checkmark$/X on green/brown background indicates an active/silent mode. For some resonances, the orientation of the oscillating electric field with respect to a crystallographic axis determine if it is active. In this case, $\checkmark_y$ labels resonances active only for $\mathbf{E^\omega} \parallel \mathbf{Y}$. The experimental results (Exp.) are compared to the polarization matrix elements derived from the different spin-polarization coupling terms (for details see the text). In the model calculations, we choose $\mathbf{X} \parallel \mathbf{q}$, thus, we include the experimental selection rules for both $\mathbf{E}^\omega$ $\parallel$ $\mathbf{Y}$ and $\mathbf{E}^\omega$ $\parallel$ $\mathbf{Z} \times \mathbf{q}$ in the same table as the theoretical results.}
\label{table:selectionrules_EZxq}
\end{table*}

Since we measured the spectra in high magnetic fields first and the $\mathbf{q}$ vector remains nearly perpendicular to the field even as $B\rightarrow0$ \cite{Bordacs2018}, the strengths of the modes for a given polarization are different for $\mathbf{B} \parallel \mathbf{X}$ and $\mathbf{B} \parallel \mathbf{Y}$ even when the field is decreased to zero (compare (a) to (c) and (b) to (d) pairs in Figs.~\ref{fig_Bdcplane_Ew111}-\ref{fig_Bdcplane_BwEwinplane}). Correspondingly, the zero-field selection rules could be determined for cycloids with $\mathbf{q} \parallel \mathbf{Y}$ and $\mathbf{q} \parallel \mathbf{X}$. In general, the selection rules are similar for both directions of the $\mathbf{q}$ vector if the oscillating fields are considered in the frame rotated together with the cycloidal $\mathbf{q}$ vector. More precisely, when $\mathbf{E^\omega} \parallel \mathbf{Z}$ and $\mathbf{B^\omega} \perp \mathbf{Z}$ the absorption spectra are the same in the rows of Fig.~\ref{fig_Bdcplane_Ew111}. In this case, only the relative orientation of $\mathbf{B^\omega}$ with respect to $\mathbf{q}$ determines which modes are excited. However, when $\mathbf{E^\omega}$ is in the $\mathrm{XY}$ plane, the different relative intensities in the rows of Fig.~\ref{fig_Bdcplane_Bw111} and Fig.~\ref{fig_Bdcplane_BwEwinplane} suggest that in case of the electric fields, their orientation relative to crystallographic axes could also matter. The lower branch of the 16.3\,cm$^{-1}$ mode serves as an example since it is active for $\mathbf{E^\omega} \parallel \mathbf{Y}$ (Fig.~\ref{fig_Bdcplane_Bw111}.~(b) and (d)) but not for $\mathbf{E^\omega} \parallel \mathbf{X}$ (Fig.~\ref{fig_Bdcplane_Bw111}.~(a) and (c)). The magnetic and electric dipole selection rules determined for the different modes are summarized in Tables \ref{table:selectionrules_B} and \ref{table:selectionrules_EZ}-\ref{table:selectionrules_EZxq}. In the following, we analyse them for each mode. 

We start with one of the strongest modes, $\Psi \mathit{_{1}^{(1)}}$ at 22.2\,cm$^{-1}$ in zero field. As $B\rightarrow0$, on one hand this mode is active only for $\mathbf{B^\omega} \parallel \mathbf{X}$ and $\mathbf{E^\omega} \parallel \mathbf{Y}$ when $\mathbf{B} \parallel \mathbf{Y}$, and on the other hand, it can be excited by $\mathbf{B^\omega} \parallel \mathbf{Y}$ and $\mathbf{E^\omega} \parallel \mathbf{X}$ when $\mathbf{B} \parallel \mathbf{X}$ (see and compare Figs.~\ref{fig_Bdcplane_Ew111}-\ref{fig_Bdcplane_BwEwinplane}). Therefore, $\mathbf{B^\omega} \parallel \mathbf{q}$ and $\mathbf{E^\omega} \parallel \mathbf{Z} \times \mathbf{q}$ excite this resonance as predicted by theory for the $\Psi \mathit{_{1}^{(1)}}$ mode \cite{Fishman2013PRB}. In a finite field, it gains magnetic dipole strength if $\mathbf{B^\omega} \parallel \mathbf{Z}$ as its intensity gradually grows with the field (see Fig.~\ref{fig_Bdcplane_Bw111} (c) and (d)). Moreover, it acquires finite NDD in field as shown in Fig.~\ref{fig_Bdcplane_Bw111} (a) and (b). As NDD requires that the excitation is active for both fields, $\mathbf{E^\omega}$ and $\mathbf{B^\omega}$, this also supports that $\Psi \mathit{_{1}^{(1)}}$ becomes active for $\mathbf{B^\omega} \parallel \mathbf{Z}$. Although $\Psi \mathit{_{1}^{(1)}}$ can be excited by $\mathbf{B^\omega} \parallel \mathbf{q}$ and $\mathbf{E^\omega} \parallel \mathbf{Z} \times \mathbf{q}$, we could not resolve NDD for this light polarization (Fig.~\ref{fig_Bdcplane_BwEwinplane} (c) and (d)).

The other strong resonance, $\Psi \mathit{_{1}^{(2)}}$ at 20.3\,cm$^{-1}$ in zero field, is absent for $\mathbf{B^\omega} \parallel \mathbf{X}$ and $\mathbf{B^\omega} \parallel \mathbf{Y}$ when $\mathbf{E^\omega} \parallel \mathbf{Z}$ (see Fig.~\ref{fig_Bdcplane_Ew111}), thus, neither the in-plane $\mathbf{B^\omega}$ nor the $\mathbf{E^\omega} \parallel \mathbf{Z}$ excites this resonance. As $\Psi \mathit{_{1}^{(2)}}$ is present in all panels of Fig.~\ref{fig_Bdcplane_BwEwinplane} and it is silent for $\mathbf{B^\omega} \perp \mathbf{Z}$, both in-plane electric fields $\mathbf{E^\omega} \parallel \mathbf{X}$ and $\mathbf{E^\omega} \parallel \mathbf{Y}$ couple to this resonance. The larger oscillator strength of $\Psi \mathit{_{1}^{(2)}}$ in panel (a) and (c) of Fig.~\ref{fig_Bdcplane_Bw111} compared to panel (b) and (d) suggests stronger interaction for $\mathbf{E^\omega} \parallel \mathbf{X}$. Furthermore, this mode is active for all polarizations in Fig.~\ref{fig_Bdcplane_Bw111}, thus, $\mathbf{B^\omega} \parallel \mathbf{Z}$ may also excite it, which would agree with the theoretical expectations \cite{Fishman2013PRB}. In a finite magnetic field, $\Psi \mathit{_{1}^{(2)}}$ gains magnetic dipole character for $\mathbf{B^\omega} \parallel \mathbf{q}$ as evidenced in Fig.~\ref{fig_Bdcplane_Ew111}. While the direction of the oscillating electric field is the same, $\mathbf{E^\omega} \parallel \mathbf{Z}$ in all panels its intensity grows only when $\mathbf{B^\omega} \parallel \mathbf{q}$ (panel (c) and (d)), but it remains silent for $\mathbf{B^\omega} \parallel \mathbf{Z} \times \mathbf{q}$ (panel (a) and (b)). Correspondingly, finite NDD arises in finite fields for this simultaneously electric and magnetic dipole active resonance when $\mathbf{B^\omega} \parallel \mathbf{q}$ and $\mathbf{E^\omega} \perp \mathbf{Z}$ (see Fig.~\ref{fig_Bdcplane_BwEwinplane} (c) and (d)). We note that NDD is also observed for the $\Psi \mathit{_{1}^{(2)}}$ mode in panel (a) and (b) of Fig.~\ref{fig_Bdcplane_BwEwinplane}, where it should not be excited by the magnetic component of the radiation, $\mathbf{B^\omega} \parallel \mathbf{Z} \times \mathbf{q}$. For this light polarization the origin of NDD is unclear.

In zero field, the modes that are degenerate at 16.3\,cm$^{-1}$ are absent in Fig.~\ref{fig_Bdcplane_Bw111}, thus, they are silent for $\mathbf{B^\omega} \parallel \mathbf{Z}$, $\mathbf{E^\omega} \parallel \mathbf{X}$ and $\mathbf{E^\omega} \parallel \mathbf{Y}$. Although we could not resolve any zero-field splitting similar to previous studies \cite{Nagel2013,Kezsmarki2015, Talbayev2011PRBBiFeO}, we can extrapolate the field dependence of the intensities to zero field to deduce the corresponding selection rules. In Fig.~\ref{fig_Bdcplane_Ew111} (a) and (b), the upper branch is absent but appears in panel (c) and (d), thus, only $\mathbf{B^\omega} \parallel \mathbf{q}$ excites it. The vanishing intensity of the lower branch in panel (c) and (d) suggests that it is silent for $\mathbf{B^\omega} \parallel \mathbf{q}$ in zero field. However, it has finite strength in panels (a) and (b), thus, it should be active for $\mathbf{B^\omega} \parallel \mathbf{Z} \times \mathbf{q}$. These selection rules also agree with the spectra displayed in Fig.~\ref{fig_Bdcplane_BwEwinplane}. Based on these findings, we ascribe the $\mathbf{B^\omega} \parallel \mathbf{q}$ active upper branch to the $\Psi \mathit{_{0}}$ mode, whereas the lower branch excited by $\mathbf{B^\omega} \parallel \mathbf{Z} \times \mathbf{q}$ corresponds to the $\Phi \mathit{_{1}^{(1)}}$ spin-wave. This identification of the modes contradicts previous THz studies, where the light propagation was along the pseudo-cubic [001] direction \cite{Nagel2013, Kezsmarki2015}. However, our experiments performed on the principal cuts provide clear selection rules supporting the present assignment. 

In Fig.~\ref{fig_Bdcplane_Ew111} (a) and (b), the intensity of the $\Phi \mathit{_{1}^{(1)}}$ mode becomes negligible for negative fields between -6\,T and -12\,T evidencing strong NDD. This implies that in finite fields, $\Phi \mathit{_{1}^{(1)}}$ is active for both $\mathbf{B^\omega} \parallel \mathbf{Z} \times \mathbf{q}$ and $\mathbf{E^\omega} \parallel \mathbf{Z}$. Phenomenologically, the appearance of the NDD can be attributed to the fact that the light propagation is parallel to the cross product of the static ferroelectric polarization, $\mathbf{P}$ and the field induced magnetization, $\mathbf{M}$ which can be viewed as a toroidal moment, $\mathbf{T} = \mathbf{P} \times \mathbf{M}$ \cite{TOKURA20071145, Szaller2013}. We note that in this case the light beam propagates along the $\mathbf{q}$ vector of the cycloidal order. 
  
The $\Phi \mathit{_{2}^{(1,2)}}$ modes at about 27.3\,cm$^{-1}$ in zero field are present for most light polarizations, so we can not determine specific selection rules. Based on Fig.~\ref{fig_Bdcplane_Ew111}, we expect that $\mathbf{E^\omega} \parallel \mathbf{Z}$ or in-plane $\mathbf{B^\omega}$ fields can excite $\Phi \mathit{_{2}^{(1,2)}}$. The finite NDD observed in Fig.~\ref{fig_Bdcplane_Ew111} (a) and (b) suggest, that $\Phi \mathit{_{2}^{(1,2)}}$ are active for $\mathbf{E^\omega} \parallel \mathbf{Z}$ and $\mathbf{B^\omega} \parallel \mathbf{Z} \times \mathbf{q}$. The intensity of these modes fades away in high fields approaching the cycloidal to canted-antiferromagnetic transition as shown in Figs.~\ref{fig_Bdcplane_Ew111}-\ref{fig_Bdcplane_BwEwinplane}.

The $\Psi \mathit{_{2}^{(1,2)}}$ modes are silent in zero field. They remain silent for $\mathbf{E^\omega} \parallel \mathbf{Z}$ and $\mathbf{B^\omega} \parallel \mathbf{Z} \times \mathbf{q}$ even in high field, but gain magnetic dipole strength if $\mathbf{B^\omega} \parallel \mathbf{q}$ (Fig.~\ref{fig_Bdcplane_Ew111} (a)-(d)). Since the two modes have finite intensity in Fig.~\ref{fig_Bdcplane_BwEwinplane} (a) and (b), $\mathbf{E^\omega} \parallel \mathbf{q}$ must couple to these resonances. They are present for all polarization states displayed in Fig.~\ref{fig_Bdcplane_Bw111}, thus, $\mathbf{B^\omega} \parallel \mathbf{Z}$ or $\mathbf{E^\omega} \perp \mathbf{Z} $ fields excite them. The mode pair $\Phi \mathit{_{3}^{(1,2)}}$ also appears only in finite fields. It has some intensity for almost all light polarizations above 14\,T. For these modes, $\Psi \mathit{_{2}^{(1,2)}}$ and $\Phi \mathit{_{3}^{(1,2)}}$, we could not unambiguously detect NDD due to their weak absorption.

\section{\label{sec:resultsmemeffect}Magnetic field history dependence of the absorption spectra}

\begin{figure}[th!]
\centering
\includegraphics[width=\columnwidth]{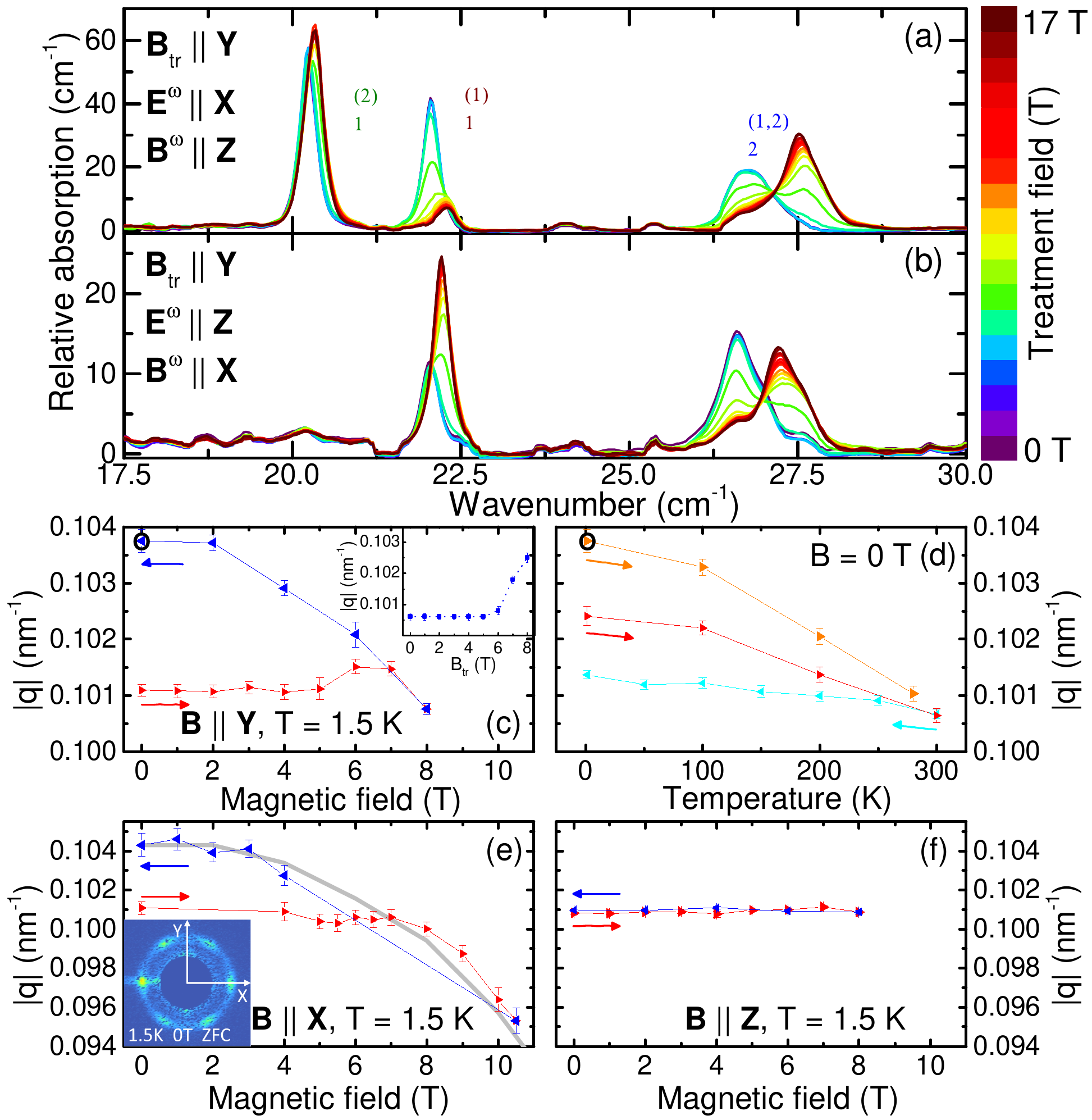}
\caption{(a)-(b): Zero-field THz absorption spectra measured at 3\,K after the application of fields $\mathbf{B}_{tr} \parallel \mathbf{Y}$ with magnitude continously increasing to 17\,T. The sample was heated above the Néel-temperature and ZFC to 3\,K before each sequence. The colour bar shows the highest treatment field, $\mathbf{B}_{tr}$ preceding each zero-field measurement. The magnetic field dependence of the $\mathbf{q}$ vector lengths, $|\mathbf{q}|$ measured with SANS at 2\,K for $\mathbf{B} \parallel \mathbf{Y}$, $\mathbf{B} \parallel \mathbf{X}$ and $\mathbf{B} \parallel \mathbf{Z}$ are plotted in panel (c), (e) and (f), respectively. A typical SANS pattern is shown in the inset of panel (e). (d) Temperature dependence of the $\mathbf{q}$ vector lengths for ZFC (light blue) and field treated (orange and red, in 8\,T when $\mathbf{B} \parallel \mathbf{Y}$) samples in zero field.}
\label{fig_memeffect}
\end{figure}

We observed that the absorption spectra measured in zero field depend on the history of the sample, in agreement with Ref.~\cite{Nagel2013}. In order to systematically study the magnetic field induced changes, we measured the zero-field spectra in the ZFC state first and then in each field applied in sequence $\mathbf{B} \in \{0, 1, 0, 2 \ldots 0, 17, 0 \}$. All three high-symmetry cuts were measured in Faraday configuration with light polarized along the two high-symmetry directions of the crystal planes, resulting in six series of spectra. We detected significant changes in the \emph{zero-field} spectra after exposing the sample to treatment field $\mathbf{B}_{tr}\parallel \mathbf{Y}$ (see Fig.~\ref{fig_memeffect} (a),(b)) and $\mathbf{B}_{tr} \parallel \mathbf{X}$ (not shown), whereas we observed negligible differences for $\mathbf{B}_{tr} \parallel \mathbf{Z}$ applied up to 17\,T.

We attribute the changes in strength of the absorption to the rearrangement of the cycloidal $\mathbf{q}$ vectors induced by in-plane \textbf{B}. Previous SANS experiments show that in-plane fields change the nearly equal population of the cyclodial domains in the ZFC state and rotate the $\mathbf{q}$ vectors in the $\mathrm{XY}$ plane to be perpendicular to the field \cite{Bordacs2018}. Moreover, this rearrangement of the $\mathbf{q}$ vectors persists after the removal of the field due to pinning of the cycloidal $\mathbf{q}$ vector on magnetic disorder. This, combined with our finding, that mostly the orientation of the oscillating fields with respect to the cycloidal $\mathbf{q}$ vector determines the THz selection rules, explains the field history dependence of the absorption strength.

In the following, we discuss the above scenario in details by focusing on the three strongest modes. In case of spectra plotted in Fig.~\ref{fig_memeffect} (a) $\mathbf{E^\omega} \perp \mathbf{Z}$, thus, both $\mathbf{E^\omega} \parallel \mathbf{q}$ and $\mathbf{E^\omega} \parallel\mathbf{Z} \times \mathbf{q}$ can excite spin-waves in the ZFC state with $\mathbf{q}$ vectors spanning 120$^\circ$ in the $\mathrm{XY}$ plane. After the application of $\mathbf{B}_{tr} \parallel \mathbf{Y}$ only cycloids with $\mathbf{q} \parallel \mathbf{X}$ remain, thus, for $\mathbf{E^\omega} \parallel \mathbf{X} \parallel \mathbf{q}$ the intensity of modes active only for $\mathbf{E^\omega} \parallel \mathbf{Z} \times \mathbf{q}$ should vanish and the absorption of modes excited only by $\mathbf{E^\omega} \parallel \mathbf{q}$ should be doubled. We expect no change in the oscillator strength of the modes active for $\mathbf{B^\omega} \parallel \mathbf{Z}$ since the relative orientation of the cycloids and the oscillating field does not change during the field treatment. In Fig.~\ref{fig_memeffect} (a), the intensity of $\Psi \mathit{_{1}^{(2)}}$ does not change significantly after the application of the field, which means that i) either $\mathbf{B^\omega} \parallel\mathbf{Z}$ excites it, or ii) both in-plane $\mathbf{E^\omega}$, or iii) the combination of the two cases. The intensity of $\Psi \mathit{_{1}^{(1)}}$ drops nearly to zero after the field treatment, which implies that it must be excited by $\mathbf{E^\omega} \parallel \mathbf{Z} \times \mathbf{q}$, and not by $\mathbf{E^\omega} \parallel \mathbf{q}$ or $\mathbf{B^\omega} \parallel \mathbf{Z}$. The degenerate $\Phi \mathit{_{2}^{(1,2)}}$ mode pair gains intensity, which suggests that $\mathbf{E^\omega} \parallel \mathbf{q}$ excites these spin-waves. However, the strength does not double, therefore, either $\mathbf{B^\omega} \parallel \mathbf{Z}$ or $\mathbf{E^\omega} \parallel \mathbf{Z} \times \mathbf{q}$ also has to excite $\Phi \mathit{_{2}^{(1,2)}}$.

In Fig.~\ref{fig_memeffect}~(b), the $\Psi \mathit{_{1}^{(2)}}$ excitation is absent even before the field treatment, i.e.~it is silent for $\mathbf{E^\omega} \parallel \mathbf{Z}$, $\mathbf{B^\omega} \parallel \mathbf{q}$ and $\mathbf{B^\omega} \parallel \mathbf{Z} \times \mathbf{q}$. The intensity of the $\Psi \mathit{_{1}^{(1)}}$ mode is doubled implying that $\mathbf{B^\omega} \parallel \mathbf{q}$ excites it, but $\mathbf{E^\omega} \parallel \mathbf{Z}$ and $\mathbf{B^\omega} \parallel \mathbf{Z} \times \mathbf{q}$ do not. The strength of $\Phi \mathit{_{2}^{(1,2)}}$ does not change for this light polarization allowing three possibilities: i) either $\mathbf{E^\omega} \parallel \mathbf{Z}$ excites it, or ii) both in-plane $\mathbf{B^\omega}$ couple to it with similar strength, or iii) the combination of these options. All the above conclusions deduced from the field history dependence are in agreement with the selection rules derived in Sec.~\ref{sec:resultsinplaneB}. In addition, these field-treatment experiments allowed us to obtain additional rules for the $\Phi \mathit{_{2}^{(1,2)}}$ as discussed above.

Besides the intensity of the absorption peaks, the field treatment also modifies the mode frequencies. The frequency of $\Psi \mathit{_{1}^{(2)}}$ and $\Psi \mathit{_{1}^{(1)}}$ does not change much but $\Phi \mathit{_{2}^{(1,2)}}$ modes frequency is altered considerably as it changes from 26.2\,cm$^{-1}$ to 27.3\,cm$^{-1}$. While increasing the magnitude of $\mathbf{B}_{tr}$ above 5\,T, the absorption peak observed in the ZFC state at 26.2\,cm$^{-1}$ decreases and is strongly suppressed for higher $\mathbf{B}_{tr}$ as shown in Fig.\ref{fig_memeffect}~(a)-(b). Simultaneously, another peak appears at 27.3\,cm$^{-1}$ and gains the same, or even higher oscillator strength as $\mathbf{B}_{tr}$ is increased. The small increase of the integrated intensity of the two peaks can be attributed to the rotation of the $\mathbf{q}$ vectors. We emphasize, that the two well separated peaks at about 26.2\,cm$^{-1}$ and 27.3\,cm$^{-1}$ correspond to the mutual step-like change of the zero-field frequency of the $\Phi \mathit{_{2}^{(1,2)}}$ modes which remain degenerate. The frequency of this mode pair in Figs.~\ref{fig_Bdcplane_Ew111}-\ref{fig_Bdcplane_BwEwinplane} is $\sim$27.3$\pm$0.2\,cm$^{-1}$, while the field treatment modifies it by $\sim$1\,cm$^{-1}$.

Using SANS, we found that the average length of the cycloidal $\mathbf{q}$ vectors, $|\mathbf{q}|$ also depends on the magnetic field history. Fig.~\ref{fig_memeffect}~(c), (e) and (f) show the magnetic field dependence of $|\mathbf{q}|$ for $\mathbf{B}_{tr} \parallel \mathbf{Y}$, $\parallel \mathbf{X}$ and $\parallel \mathbf{Z}$, respectively. SANS data was collected starting from the ZFC state in increasing (red symbols) and then in decreasing (blue symbols) fields. In these panels, the length of the $\mathbf{q}$ vector is normalized to the value measured after ZFC at 2\,K, which is about q$_0$=0.0101\,\AA$^{-1}$, but showed about $\sim$1.5\% variation between the different experiments performed on different instruments and at different times. In the temperature-dependent plot presented in Fig.~\ref{fig_memeffect}~(d) the orange curve was measured after the experiment presented in Fig.~\ref{fig_memeffect}~(c), thus, in these panels the same scaling factor was used for the normalization. After heating the sample above the Néel temperature and then cooling it in zero field, we exposed the sample to $\mathbf{B} \parallel \mathbf{Y}$ at low temperature again and measured $|\mathbf{q}|$ upon warming to and cooling from 300\,K, which are shown in red and cyan in Fig.~\ref{fig_memeffect}~(d), respectively.

By applying in-plane magnetic fields, $|\mathbf{q}|$ changes only above $\sim$5\,T and it decreases in high fields as displayed in Fig.~\ref{fig_memeffect}~(e). While decreasing the external magnetic field, the cycloid pitch does not relax back to the ZFC value, but it shrinks further corresponding to an increase in $|\mathbf{q}|$. A detailed sequence similar to the one used in the THz experiments ($\mathbf{B}_{tr} \in \{0, 1, 0, 2 \ldots 0,8, 0 \}$) revealed, that the zero field value of $|\mathbf{q}|$ starts to change above 5\,T, and with increasing magnitude of $\mathbf{B}_{tr}$, it continuously grows up to approximately 103\% of the ZFC length. After the sample is warmed to room temperature and cooled back again, $|\mathbf{q}|$ changes back to the ZFC value (Fig.~\ref{fig_memeffect}~(d)) even though the field-induced orientation of the cycloids remains intact. By applying $\mathbf{B} \parallel \mathbf{Z}$, $|\mathbf{q}|$ does not change up to the highest available field of 8\,T, Fig.~\ref{fig_memeffect}~(f).

Our results suggest that the change in the zero-field excitation energies of the spin-waves is caused by the field-induced change of $|\mathbf{q}|$. Although in the THz experiments we detected $\Phi \mathit{_{2}^{(1,2)}}$ modes in two well-defined positions before and after the application of the field, SANS data indicates rather continuous change of $|\mathbf{q}|$. This conflict can be resolved by noting that the instrumental resolution of the SANS experiments was $\Delta q\sim 0.0011$ \AA$^{-1}$, which does not allow us to resolve features smaller then 10\% in $|\mathbf{q}|$. We assume that after ZFC to 2\,K the cycloid is in a metastable state with a longer wavelength. The shorter $|\mathbf{q}|$ results in lower excitation energies if the exchange parameters remain the same as the size of the magnetic Brillouin zone is reduced. The application of a magnetic field forces the cycloidal wavelength to expand and thus the wavefronts to move, may allow the cycloids to relax and to reach the stable configuration with longer $|\mathbf{q}|$. This hypothesis is supported by the facts that i) at 300\,K $|\mathbf{q}|$ is the same when the sample is ZFC or warmed from 2\,K after the field cycling, ii) $|\mathbf{q}|$ is the same at 300\,K before and after the application of a magnetic field.

\section{\label{sec:resultsBz}THz spectra for magnetic fields applied along Z}

\begin{figure*}[th!]
\centering
\includegraphics[width=\textwidth]{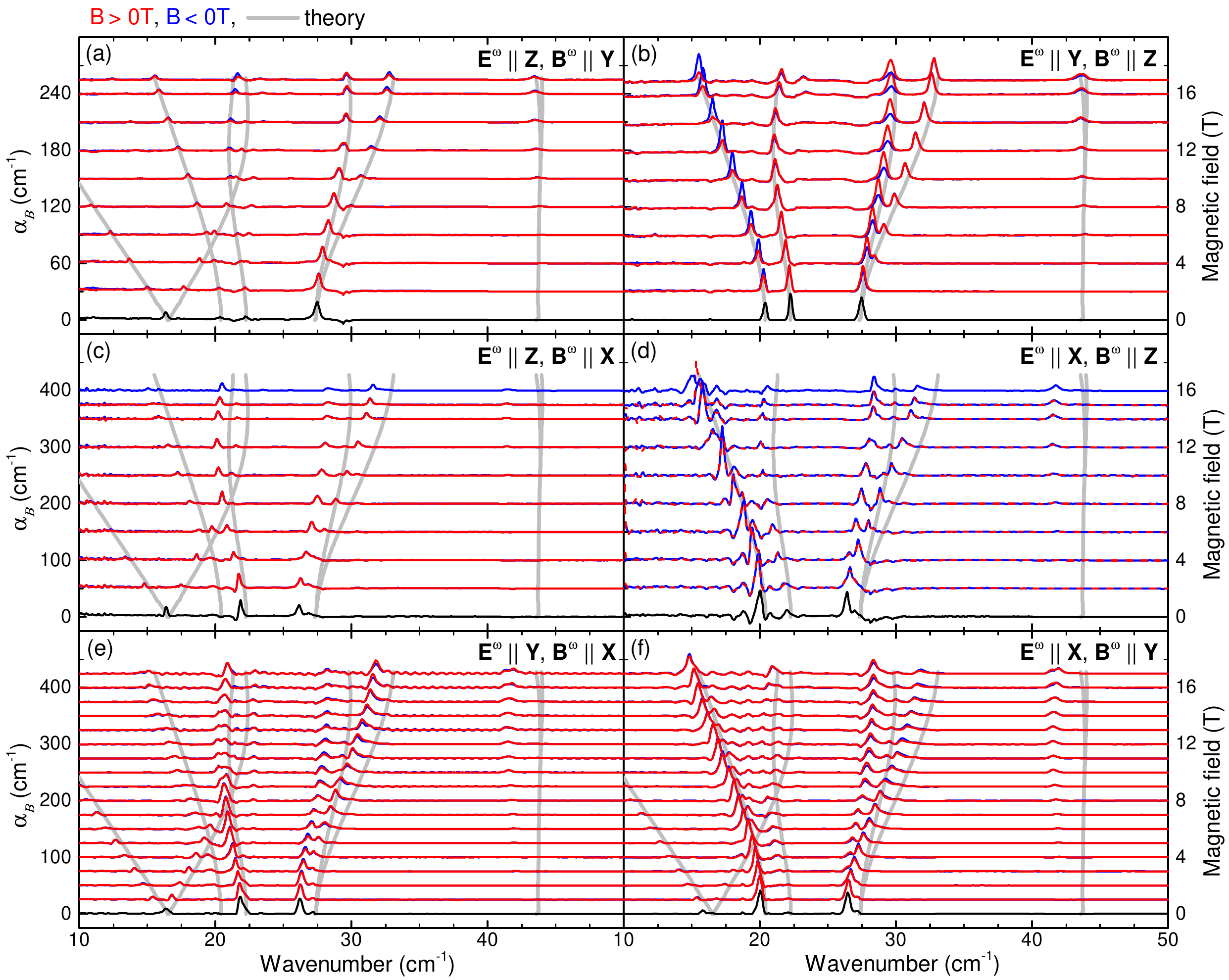}
\caption{Magnetic field dependence of the THz absorption spectra of BiFeO$_3$ at 4\,K. The magnetic field is applied along $\mathbf{Z}$. The spectra are plotted for positive and negative fields in red and blue, respectively, with a vertical offset proportional to the magnitude of the field. In panel (d), the spectra measured in negative fields are plotted with dashed blue lines to highlight that they are identical within the accuracy of the experiment to that of obtained in positive fields. Grey lines show the magnetic field dependence of spin-wave frequencies deduced from linear spin-wave theory.}
\label{fig_Bdc111_measured_data}
\end{figure*}

Spectra measured in $\mathbf{B} \parallel \mathbf{Z}$ with light polarized along the three principal directions are shown in Fig.~\ref{fig_Bdc111_measured_data}. Panels in the same row correspond to the experiments performed on the same crystal cut using two orthogonal light polarizations. Spectra measured in positive and negative fields are plotted in red and blue, respectively. In panel (d), the spectra measured in positive fields are plotted with dashed red lines to illustrate that they are identical within the accuracy of the experiment to that obtained in negative fields (blue lines). However, we observed strong NDD in panel (b) for modes $\Psi \mathit{_{1}^{2}}$ and $\Phi \mathit{_2^1}$, and a smaller effect in panel (a) for modes $\Psi \mathit{_{1}^{1}}$ and $\Phi \mathit{_2^2}$.

In the magnetic field dependence of the resonance energies a shift is observed in Fig.~\ref{fig_Bdc111_measured_data} (c)-(f) compared to panels (a)-(b). In case of Fig.~\ref{fig_Bdc111_measured_data} (a)-(b) the $\mathrm{YZ}$ face sample was exposed to in-plane fields prior to the experiments whereas the spectra in panel (c)-(f) are measured on ZFC samples. As discussed in Sec.~\ref{sec:resultsmemeffect} $|\mathbf{q}|$ is altered by the application of in-plane magnetic fields, however, it does not change for $\mathbf{B} \parallel \mathbf{Z}$ in the applied field range (see Fig.~\ref{fig_memeffect}.~(f)). Therefore, we can ascribe the observed discrepancy to the 3\% shorter $|\mathbf{q}|$ in panels (c)-(f) compared to panels (a)-(b), which alters the spin-wave frequencies.

It is difficult to deduce clear selection rules in $\mathbf{B} \parallel \mathbf{Z}$ for the following reasons. First, the field $\mathbf{B} \parallel \mathbf{Z}$ does not establish a single cycloidal domain state, therefore, the domain distribution is not well defined. Second, the spin-wave modes are active for most of the light polarizations. Since $\Phi \mathit{_1^1}$ and $\Psi \mathit{_0}$ are clearly absent in panel (b) and (d), they must be forbidden for $\mathbf{E^\omega} \parallel \mathbf{X}$, $\mathbf{E^\omega} \parallel \mathbf{Y}$, and $\mathbf{B^\omega} \parallel \mathbf{Z}$. However, these resonances are present in panel (e) and (f), thus, they should be active for oscillating in-plane magnetic fields $\mathbf{B^\omega} \parallel \mathbf{X}$ and $\mathbf{B^\omega} \parallel \mathbf{Y}$. The finite NDD observed in Fig.~\ref{fig_Bdc111_measured_data} (a) and (b) provides further insights: Modes $\Psi \mathit{_{1}^{2}}$ and $\Phi \mathit{_2^1}$ must couple to both $\mathbf{E^\omega} \parallel \mathbf{Y}$, and $\mathbf{B^\omega} \parallel \mathbf{Z}$ (see Fig.~\ref{fig_Bdc111_measured_data} (b)). Similarly, modes $\Psi \mathit{_{1}^{1}}$ and $\Phi \mathit{_2^2}$ should be active for $\mathbf{E^\omega} \parallel \mathbf{Z}$ and $\mathbf{B^\omega} \parallel \mathbf{Y}$ (Fig.~\ref{fig_Bdc111_measured_data} (a)). The selection rules for the higher-energy weak resonances are ambiguous.

\begin{figure}[h!]
\centering
\includegraphics[width=\columnwidth]{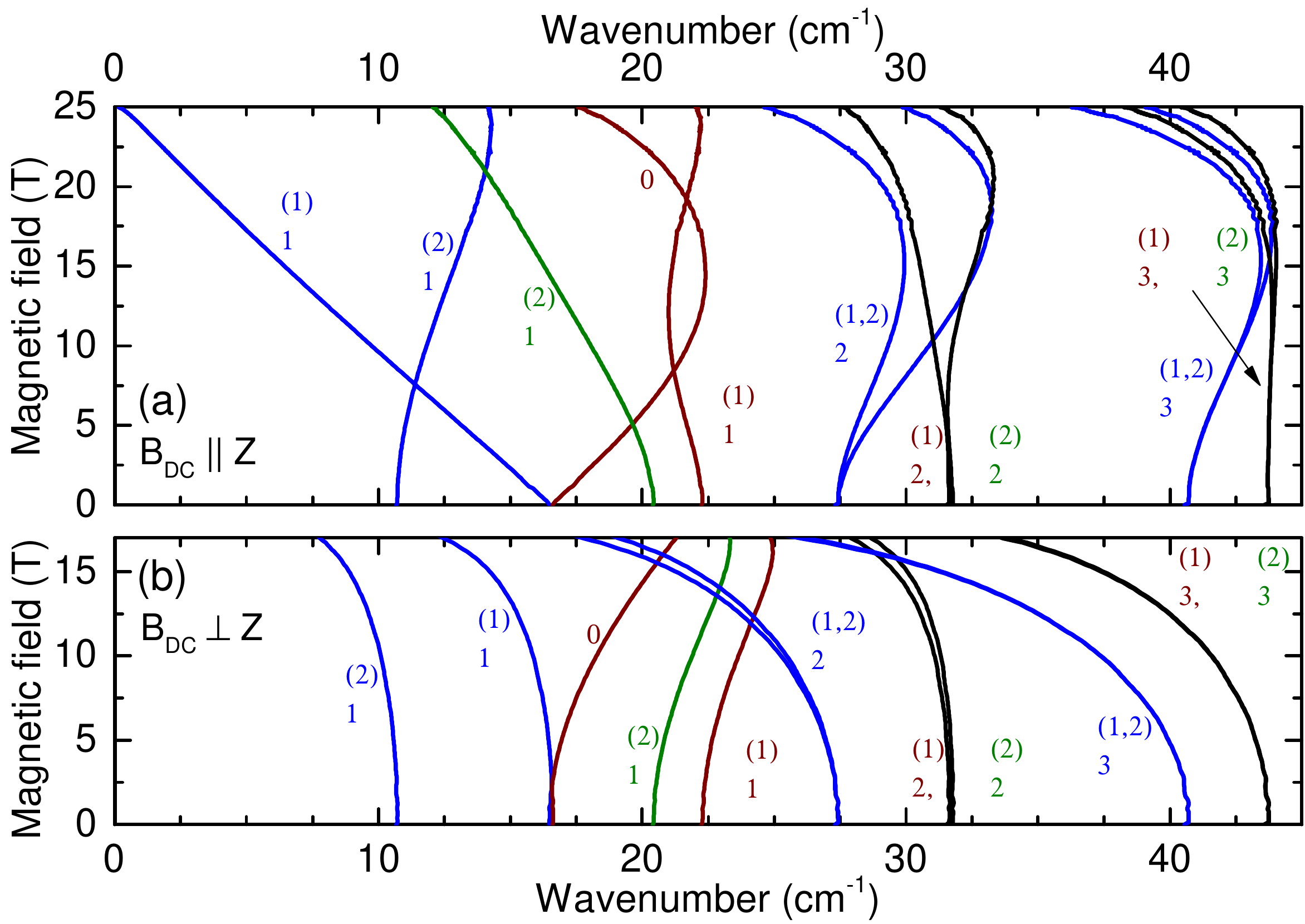}
\caption{(a), (b) Magnetic field dependence of spin-wave frequencies in BiFeO$_3$ for $\mathbf{B} \parallel \mathbf{Z}$ and $\mathbf{B} \parallel \mathbf{Y}$, respectively, as deduced from linear spin-wave theory. There are three types of spin-wave modes corresponding to spin rotation about the $\mathbf{Z}$ ($\Psi \mathit{_{n}^{2}}$, green), $\mathbf{X} \parallel \mathbf{q}$ ($\Psi \mathit{_0}$ and $\Psi \mathit{_{n}^{1}}$, brown) and $\mathbf{Y} \parallel \mathbf{Z} \times \mathbf{q}$ ($\Phi \mathit{_n}$, blue) axes.}
\label{fig_Randy_theory}
\end{figure}

\section{\label{sec:Theory}Spin-wave theory}

We modelled the spin excitations based on a linear spin-wave theory developed in Refs.~\onlinecite{Fishman2013PRB, Fishman2013}. We calculated the magnetic field dependence of the spin-wave energies for fields applied along the high-symmetry directions, $\mathbf{B} \parallel \mathbf{Y}$ and $\mathbf{B} \parallel \mathbf{Z}$ while keeping $\mathbf{q} \parallel \mathbf{X}$. The results are displayed in Fig.~\ref{fig_Randy_theory}. We found a good agreement with the experiments for $\mathbf{B} \parallel \mathbf{Y} \parallel \mathbf{Z} \times \mathbf{q}$ (see grey curves in Figs.~\ref{fig_Bdcplane_Ew111}-\ref{fig_Bdcplane_BwEwinplane}) using the same coupling constants as in Ref.~\onlinecite{Kezsmarki2015}. The theoretical results for $\mathbf{B} \parallel \mathbf{Z}$ also describe the magnetic field dependence of the observed resonances (Fig.~\ref{fig_Bdc111_measured_data}) except the frequency  shift discussed in Sec.~\ref{sec:resultsBz}. 

In order to reproduce the selection rules deduced above for in-plane fields, we calculated the magnetic, $\langle 0|M|n \rangle$ and electric dipole $\langle 0|P|n \rangle$ matrix elements between the ground state $|0 \rangle$ and the excited state $|n \rangle$ for $\mathbf{B} \parallel \mathbf{Y} \parallel \mathbf{Z} \times \mathbf{q}$. As shown in Table~\ref{table:selectionrules_B}, our model describes well the magnetic dipole selection rules implying that our spin-wave theory correctly captures the spin dynamics. In zero field, we identified three types of spin-wave modes corresponding to spin rotation around the $\mathbf{Z}$ (modes $\Psi \mathit{_{n}^{(2)}}$), $\mathbf{X} \parallel \mathbf{q}$ ($\Psi \mathit{_0}$ and $\Psi \mathit{_{n}^{(1)}}$) and $\mathbf{Y} \parallel \mathbf{Z} \times \mathbf{q}$ ($\Phi \mathit{_n}$) axes. Next, we considered three basic mechanisms in the calculation of the spin induced polarization: the spin-current (SC), the magnetostriction (MS) and the single-ion mechanism (ANI) \cite{Fishman2015}. The definition of these polarization terms are summarized in the Appendix. The electric dipole selection rules are presented in Table \ref{table:selectionrules_Eq}, \ref{table:selectionrules_EZxq} and \ref{table:selectionrules_EZ} for oscillating electric field along $\mathbf{X}$, $\mathbf{Y}$ and $\mathbf{Z}$, respectively. Solely based on the presence (or absence) of the electric dipole strength we cannot uniquely identify a mechanism responsible for all selection rules. Instead, polarization terms SC(1)$_X$ and ANI(2)$_X$ are consistent with the experimental results for $\mathbf{E^\omega} \parallel \mathbf{X}$ (see Table \ref{table:selectionrules_Eq}) though both of them predicts a negligible dipole moment for $\Psi \mathit{_0}$ in finite fields. In case of $\mathbf{E^\omega} \parallel \mathbf{Y}$, SC(1)$_Z$ and ANI(1)$_Y$ are generating polarization along $\mathbf{Y}$, in agreement with the experiment (see Table~\ref{table:selectionrules_EZxq}). We note, that even SC(2)$_Y$, MS(2)$_Y$ and ANI(4)$_Y$ might explain the observed selection rules for $\mathbf{E^\omega} \parallel \mathbf{Y}$ as they predict only a small contribution to the absorption of mode $\Psi \mathit{_0}$. Finally, polarization terms SC(2)$_Z$, MS(1)$_Z$ and ANI(3)$_Z$ are all compatible with the experimental results for $\mathbf{E^\omega} \parallel \mathbf{Z}$ (see Table \ref{table:selectionrules_EZ}).

\begin{figure}[h!]
\centering
\includegraphics[width=\columnwidth]{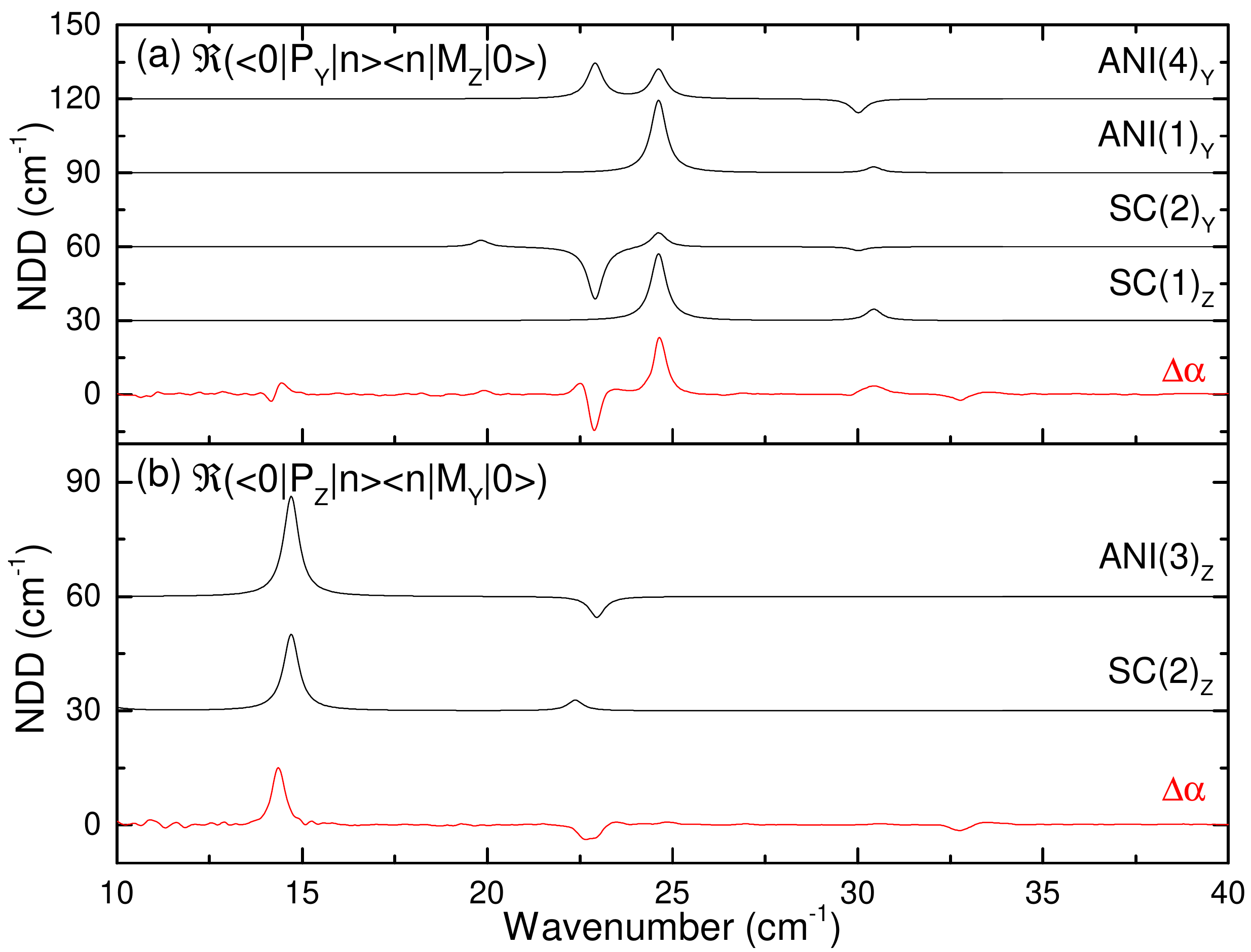}
\caption{(a)-(b) The NDD spectra calculated in 14\,T for light polarization $\mathbf{E^\omega} \parallel \mathbf{Y}$, $\mathbf{B^\omega} \parallel \mathbf{Z}$ and $\mathbf{E^\omega} \parallel \mathbf{Z}$, $\mathbf{B^\omega} \parallel \mathbf{Y}$, respectively. The theoretical results are compared to the difference of the spectra measured in positive and negative fields, $\Delta \alpha = \alpha_B (+B)- \alpha_B (-B)$.}
\label{fig_DDmechs}
\end{figure}

The analysis of the NDD can provide further information on the dynamic spin-polarization coupling. Since NDD is proportional to the real part of the product of the electric and magnetic dipole matrix elements, $\Delta \alpha \propto \Re(\langle 0|P|n \rangle \langle n|M|0 \rangle)$ \cite{Kezsmarki2011,Bordacs2012}, the relative phase of the oscillating dipoles becomes important. The theory predicts finite NDD only for two cases, $\mathbf{E^\omega} \parallel \mathbf{Y}$, $\mathbf{B^\omega} \parallel \mathbf{Z}$ and $\mathbf{E^\omega} \parallel \mathbf{Z}$, $\mathbf{B^\omega} \parallel \mathbf{Y}$, as shown in Fig.~\ref{fig_DDmechs} (a) and (b), respectively. 

Among the spin induced polarization terms only SC(2)$_Y$ can reproduce the opposite sign of the NDD signal observed for mode $\Psi \mathit{_{2}^{(1)}}$ and $\Psi \mathit{_{1}^{(1)}}$ as shown in Fig.~\ref{fig_DDmechs} (a). Beside this, either SC(1)$_Z$ or ANI(1)$_Y$ or both could contribute to P$_Y$ in order to describe the stronger NDD of $\Psi \mathit{_{1}^{(1)}}$. However, we can discard mechanisms ANI(4)$_Y$ and MS(2)$_Y$ as they respectively predict opposite NDD for modes $\Psi \mathit{_{1}^{(1)}}$ and $\Psi \mathit{_{2}^{(1,2)}}$, and vanishing NDD.

For the orthogonal light polarization (see Fig.~\ref{fig_DDmechs} (b)), only ANI(3)$_Y$ captures the sign difference of the NDD for modes $\Phi \mathit{_{1}^{(1)}}$ and $\Phi \mathit{_{2}^{(1,2)}}$, thus, it is essential to describe P$_Z$. SC(2)$_Z$ may give a minor contribution to P$_Z$ if any, whereas, MS(1)$_Z$ does not generate NDD at all. 

Finally, we note that NDD is vanishing for both orthogonal light polarizations in the Faraday geometry in agreement with the experiments. However, we did observed finite NDD when both fields oscillate in the Z plane (see Fig.~\ref{fig_Bdcplane_BwEwinplane}) that neither of the polarization mechanisms can reproduce.

\section{\label{sec:conclusion}Conclusions}

In conclusion, we systematically studied the magnetic field dependence of the spin-wave excitations of BiFeO$_3$ in high-symmetry cuts of a ferroelectric mono-domain single crystal by THz spectroscopy. In fields applied in the plane normal to the ferroelectric polarization along $\mathbf{Z}$, the resonance frequencies follow the same field dependence irrespective of the field direction, $\mathbf{B} \parallel \mathbf{X}$ or $\mathbf{B} \parallel \mathbf{Y}$. We attributed this to the reorientation of cycloidal $\mathbf{q}$ vectors perpendicular to the in-plane magnetic field. We found that the selection rules are mostly determined by the relative orientation of the oscillating fields with respect to the cycloidal $\mathbf{q}$ vectors and only the electric dipole coupling of a few modes may be sensitive to the orientation of the crystallographic axes. For all three magnetic field directions we detected strong NDD. The observed field history dependence of the absorption spectra is explained by the interplay of the hysteretic magnetic field induced reorientation of the $\mathbf{q}$ vectors and by the changes in the $\mathbf{q}$ vector length. 

We found that our linear spin-wave theory describes the spin dynamics correctly: the field dependence of the spin-wave frequencies as well as the magnetic selection rules. From the observed electric dipole selection rules and from the directional dichroism spectra, we come to the conclusion that none of the spin-induced-polarization mechanism alone can be responsible for the dynamic magnetoelectric effect. The simplest form of the spin-current interaction \citep{Katsura2008} cannot explain the presence of a dynamic spin-induced polarization by $\mathbf{E^\omega} \parallel \mathbf{q}$. At least a combination of two mechanisms, spin-current and anisotropy, is necessary to describe the experimental NDD. Our results demonstrate that THz absorption and directional dichroism spectroscopy combined with theoretical calculations may allow to distinguish between the possible magnetoelectric coupling terms. Moreover, understanding the microscopic origin of the NDD may promote its application in microwave or THz communication \cite{Kezsmarki2015}.

\begin{acknowledgments}
This research was supported by the Estonian Ministry of Education and Research grants IUT23-3, PRG736, by the European Regional Development Fund Project No.~TK134, by the bilateral program of the Estonian and Hungarian Academies of Sciences under the Contract NMK2018-47, by the Hungarian National Research, Development and Innovation Office – NKFIH grants ANN 122879 and FK 135003. The research reported in this paper and carried out at the BME has been supported by the NRDI Fund (TKP2020 IES, Grant No.~BME-IE-NAT) based on the charter of bolster issued by the NRDI Office under the auspices of the Ministry for Innovation and Technology. D.~Sz.~acknowledges the support of the Austrian Science Fund (FWF) [I 2816-N27, TAI 334-N] and that of the Austrian Agency for International Cooperation in Education and Research [WTZ HU 08/2020]. R.~S.~F.~acknowledges support by the U.S.~Department of Energy, Office of Basic Energy Sciences, Materials Sciences and Engineering Division. The measurement data from ILL are publicly available under the ILL doi:10.5291/ILL-DATA.INTER-338.
\end{acknowledgments}

\section*{\label{sec:Appendix}Appendix: Spin induced polarization terms}

In this appendix, we list the possible spin-polarization coupling terms relevant for BiFeO$_3$ following Ref.~\onlinecite{Fishman2015}.

Among the spin-current (SC) terms SC(1) corresponding to the homogeneous DM interaction is defined as:
\begin{equation}
P^{{\rm SC (1)}}_{\alpha} \propto \sum_{\beta} \lambda^{(1)}_{\alpha \beta} \Bigl\{ \frac{1}{N} \, \sum_{\langle i,j \rangle^{[100]}}  \mathbf{S}_i \times \mathbf{S}_j \Bigr\} _{\beta},
\end{equation}
where $\alpha$ and $\beta$ goes for \textbf{X}, \textbf{Y} and \textbf{Z}. The summation is for nearest neighbours along the cubic [100] direction. The SC(1)$_\alpha$ terms in Tables~\ref{table:selectionrules_EZ}-\ref{table:selectionrules_EZxq}. correspond to the sum of the spin-operator products without the multiplication with $\lambda^{(1)}_{\alpha \beta}$. $\lambda^{(1)}_{\alpha \beta}$ has some diagonal (XX, YY) and also offdiagonal (YZ, ZY) non-zero elements, resulting in four SC(1)$_\alpha$ terms in Tables~\ref{table:selectionrules_EZ}-\ref{table:selectionrules_EZxq}. The interaction SC(2) can be derived from the alternating DM interaction:
\begin{equation}
\mathbf{P}^{{\rm SC (2)}}_{\alpha} \propto \frac{1}{N} \, \sum_{\langle i,j\rangle} (-1)^{n_i} \bigl\{ \mathbf{S}_i \times \mathbf{S}_j \bigr\}.
\end{equation}
Here, the summation goes for all nearest neighbours.

THe magnetostriction (MS) can be derived from the Heisenberg exchange interaction. The following three equations define MS(1), MS(2) and MS(3), respectively.
\begin{equation}
\mathbf{P}^{\rm MS (1)} \propto
\begin{bmatrix}
     0 \\
     \lambda_y \\
     \lambda_z
\end{bmatrix}
\cdot \frac{1}{N} \, \sum_{\langle i,j\rangle^{\mathbf{u}}} \, \mathbf{S}_i \cdot \mathbf{S}_j,
\end{equation}
where \textbf{u} contains the cubic directions [100], [010] and [001].
\begin{equation}
\mathbf{P}^{\rm MS (2)} \propto \mathbf{Z} \times \Bigl\{ \frac{1}{N} \, \sum_{\langle i,j\rangle^{\mathbf{u}}}  (-1)^{n_i} \, (\mathbf{S}_i \cdot \mathbf{S}_j) \cdot \mathbf{u} \Bigr\},
\end{equation}
where \textbf{u} contains the cubic directions [100], [010] and [001].

Finally, single ion mechanisms for spin-polarization coupling derived from on-site anisotropy terms (ANI) may induce polarization as follows:
\begin{equation}
\mathbf{P}^{{\rm ANI (1)}}_{X, Y} \propto \frac{1}{4N} \sum_i  \bigl( S_{iX} \mathbf{X} + S_{iY} \mathbf{Y} \bigr) S_{iZ},
\end{equation}
\begin{equation}
\mathbf{P}^{{\rm ANI (2)}}_{X, Y} \propto \frac{1}{4N} \sum_i \Bigl\{ \bigl( ({S_{iX}})^2 - ({S_{iY}})^2 \bigr) \mathbf{Y} + 2 S_{iX}S_{iY} \mathbf{X} \Bigr\},
\end{equation} 
\begin{equation}
\mathbf{P}^{{\rm ANI (3)}}_{Z} \propto \frac{1}{4N} \, \sum_i ({S_{iZ}})^2 \mathbf{Z},
\end{equation}
\begin{equation}
\mathbf{P}^{{\rm ANI (4)}}_{X, Y} \propto -\frac{3}{N} \sum_i (-1)^{n_i} \bigl( S_{iY} \mathbf{X} - S_{iX} \mathbf{Y} \bigr) S_{iZ},
\end{equation}
\begin{equation}
\mathbf{P}^{{\rm ANI (5)}}_{X, Y} \propto \frac{1}{4N} \sum_i (-1)^{n_i} \Bigl\{ \bigl( ({S_{iY}})^2 - ({S_{iX}})^2 \bigr) \mathbf{X} - 2 S_{iY}S_{iX} \mathbf{Y} \Bigr\}.
\end{equation}


\begin{thebibliography}{41}%
\makeatletter
\providecommand \@ifxundefined [1]{%
 \@ifx{#1\undefined}
}%
\providecommand \@ifnum [1]{%
 \ifnum #1\expandafter \@firstoftwo
 \else \expandafter \@secondoftwo
 \fi
}%
\providecommand \@ifx [1]{%
 \ifx #1\expandafter \@firstoftwo
 \else \expandafter \@secondoftwo
 \fi
}%
\providecommand \natexlab [1]{#1}%
\providecommand \enquote  [1]{``#1''}%
\providecommand \bibnamefont  [1]{#1}%
\providecommand \bibfnamefont [1]{#1}%
\providecommand \citenamefont [1]{#1}%
\providecommand \href@noop [0]{\@secondoftwo}%
\providecommand \href [0]{\begingroup \@sanitize@url \@href}%
\providecommand \@href[1]{\@@startlink{#1}\@@href}%
\providecommand \@@href[1]{\endgroup#1\@@endlink}%
\providecommand \@sanitize@url [0]{\catcode `\\12\catcode `\$12\catcode
  `\&12\catcode `\#12\catcode `\^12\catcode `\_12\catcode `\%12\relax}%
\providecommand \@@startlink[1]{}%
\providecommand \@@endlink[0]{}%
\providecommand \url  [0]{\begingroup\@sanitize@url \@url }%
\providecommand \@url [1]{\endgroup\@href {#1}{\urlprefix }}%
\providecommand \urlprefix  [0]{URL }%
\providecommand \Eprint [0]{\href }%
\providecommand \doibase [0]{https://doi.org/}%
\providecommand \selectlanguage [0]{\@gobble}%
\providecommand \bibinfo  [0]{\@secondoftwo}%
\providecommand \bibfield  [0]{\@secondoftwo}%
\providecommand \translation [1]{[#1]}%
\providecommand \BibitemOpen [0]{}%
\providecommand \bibitemStop [0]{}%
\providecommand \bibitemNoStop [0]{.\EOS\space}%
\providecommand \EOS [0]{\spacefactor3000\relax}%
\providecommand \BibitemShut  [1]{\csname bibitem#1\endcsname}%
\let\auto@bib@innerbib\@empty
\bibitem [{\citenamefont {Hill}(2000)}]{hill2000there}%
  \BibitemOpen
  \bibfield  {author} {\bibinfo {author} {\bibfnamefont {N.~A.}\ \bibnamefont
  {Hill}},\ }\href {https://doi.org/10.1021/jp000114x} {\bibinfo {title} {Why
  are there so few magnetic ferroelectrics?}} (\bibinfo {year} {2000}),\
  \Eprint {https://arxiv.org/abs/https://doi.org/10.1021/jp000114x}
  {https://doi.org/10.1021/jp000114x} \BibitemShut {NoStop}%
\bibitem [{\citenamefont {Dong}\ \emph {et~al.}(2015)\citenamefont {Dong},
  \citenamefont {Liu}, \citenamefont {Cheong},\ and\ \citenamefont
  {Ren}}]{dong2015multiferroic}%
  \BibitemOpen
  \bibfield  {author} {\bibinfo {author} {\bibfnamefont {S.}~\bibnamefont
  {Dong}}, \bibinfo {author} {\bibfnamefont {J.-M.}\ \bibnamefont {Liu}},
  \bibinfo {author} {\bibfnamefont {S.-W.}\ \bibnamefont {Cheong}},\ and\
  \bibinfo {author} {\bibfnamefont {Z.}~\bibnamefont {Ren}},\ }\bibfield
  {title} {\bibinfo {title} {Multiferroic materials and magnetoelectric
  physics: symmetry, entanglement, excitation, and topology},\ }\href@noop {}
  {\bibfield  {journal} {\bibinfo  {journal} {Advances in Physics}\ }\textbf
  {\bibinfo {volume} {64}},\ \bibinfo {pages} {519} (\bibinfo {year}
  {2015})}\BibitemShut {NoStop}%
\bibitem [{\citenamefont {Manfred}\ \emph {et~al.}(2016)\citenamefont {Manfred}
  \emph {et~al.}}]{manfred2016evolution}%
  \BibitemOpen
  \bibfield  {author} {\bibinfo {author} {\bibfnamefont {F.}~\bibnamefont
  {Manfred}} \emph {et~al.},\ }\bibfield  {title} {\bibinfo {title} {The
  evolution of multiferroics},\ }\href@noop {} {\bibfield  {journal} {\bibinfo
  {journal} {Nat. Mater}\ }\textbf {\bibinfo {volume} {1}},\ \bibinfo {pages}
  {16046} (\bibinfo {year} {2016})}\BibitemShut {NoStop}%
\bibitem [{\citenamefont {Spaldin}\ and\ \citenamefont
  {Ramesh}(2019)}]{Spaldin2019}%
  \BibitemOpen
  \bibfield  {author} {\bibinfo {author} {\bibfnamefont {N.~A.}\ \bibnamefont
  {Spaldin}}\ and\ \bibinfo {author} {\bibfnamefont {R.}~\bibnamefont
  {Ramesh}},\ }\bibfield  {title} {\bibinfo {title} {Advances in
  magnetoelectric multiferroics},\ }\href
  {https://doi.org/10.1038/s41563-018-0275-2} {\bibfield  {journal} {\bibinfo
  {journal} {Nat. Mater.}\ }\textbf {\bibinfo {volume} {18}},\ \bibinfo {pages}
  {203} (\bibinfo {year} {2019})}\BibitemShut {NoStop}%
\bibitem [{\citenamefont {Moreau}\ \emph {et~al.}(1971)\citenamefont {Moreau},
  \citenamefont {Michel}, \citenamefont {Gerson},\ and\ \citenamefont
  {James}}]{Moreau1971}%
  \BibitemOpen
  \bibfield  {author} {\bibinfo {author} {\bibfnamefont {J.}~\bibnamefont
  {Moreau}}, \bibinfo {author} {\bibfnamefont {C.}~\bibnamefont {Michel}},
  \bibinfo {author} {\bibfnamefont {R.}~\bibnamefont {Gerson}},\ and\ \bibinfo
  {author} {\bibfnamefont {W.}~\bibnamefont {James}},\ }\bibfield  {title}
  {\bibinfo {title} {Ferroelectric {BiFeO$_3$} x-ray and neutron diffraction
  study},\ }\href {https://doi.org/10.1016/S0022-3697(71)80189-0} {\bibfield
  {journal} {\bibinfo  {journal} {Journal of Physics and Chemistry of Solids}\
  }\textbf {\bibinfo {volume} {32}},\ \bibinfo {pages} {1315 } (\bibinfo {year}
  {1971})}\BibitemShut {NoStop}%
\bibitem [{\citenamefont {Park}\ \emph {et~al.}(2014)\citenamefont {Park},
  \citenamefont {Le}, \citenamefont {Jeong},\ and\ \citenamefont
  {Lee}}]{Park2014review}%
  \BibitemOpen
  \bibfield  {author} {\bibinfo {author} {\bibfnamefont {J.-G.}\ \bibnamefont
  {Park}}, \bibinfo {author} {\bibfnamefont {M.~D.}\ \bibnamefont {Le}},
  \bibinfo {author} {\bibfnamefont {J.}~\bibnamefont {Jeong}},\ and\ \bibinfo
  {author} {\bibfnamefont {S.}~\bibnamefont {Lee}},\ }\bibfield  {title}
  {\bibinfo {title} {Structure and spin dynamics of multiferroic
  {BiFeO}$_{3}$},\ }\href {http://stacks.iop.org/0953-8984/26/i=43/a=433202}
  {\bibfield  {journal} {\bibinfo  {journal} {J. Phys.: Condens. Matter}\
  }\textbf {\bibinfo {volume} {26}},\ \bibinfo {pages} {433202} (\bibinfo
  {year} {2014})}\BibitemShut {NoStop}%
\bibitem [{\citenamefont {Martin}\ \emph {et~al.}(2007)\citenamefont {Martin},
  \citenamefont {Chu}, \citenamefont {Zhan}, \citenamefont {Ramesh},
  \citenamefont {Han}, \citenamefont {Wang}, \citenamefont {Warusawithana},\
  and\ \citenamefont {Schlom}}]{Martin2007}%
  \BibitemOpen
  \bibfield  {author} {\bibinfo {author} {\bibfnamefont {L.~W.}\ \bibnamefont
  {Martin}}, \bibinfo {author} {\bibfnamefont {Y.-H.}\ \bibnamefont {Chu}},
  \bibinfo {author} {\bibfnamefont {Q.}~\bibnamefont {Zhan}}, \bibinfo {author}
  {\bibfnamefont {R.}~\bibnamefont {Ramesh}}, \bibinfo {author} {\bibfnamefont
  {S.-J.}\ \bibnamefont {Han}}, \bibinfo {author} {\bibfnamefont {S.~X.}\
  \bibnamefont {Wang}}, \bibinfo {author} {\bibfnamefont {M.}~\bibnamefont
  {Warusawithana}},\ and\ \bibinfo {author} {\bibfnamefont {D.~G.}\
  \bibnamefont {Schlom}},\ }\bibfield  {title} {\bibinfo {title} {Room
  temperature exchange bias and spin valves based on
  {BiFeO${}_{3}$/SrRuO${}_{3}$/SrTiO${}_{3}$/Si} (001) heterostructures},\
  }\href {https://doi.org/10.1063/1.2801695} {\bibfield  {journal} {\bibinfo
  {journal} {Applied Physics Letters}\ }\textbf {\bibinfo {volume} {91}},\
  \bibinfo {pages} {172513} (\bibinfo {year} {2007})},\ \Eprint
  {https://arxiv.org/abs/https://doi.org/10.1063/1.2801695}
  {https://doi.org/10.1063/1.2801695} \BibitemShut {NoStop}%
\bibitem [{\citenamefont {Martin}\ \emph {et~al.}(2010)\citenamefont {Martin},
  \citenamefont {Chu},\ and\ \citenamefont {Ramesh}}]{Martin2010}%
  \BibitemOpen
  \bibfield  {author} {\bibinfo {author} {\bibfnamefont {L.}~\bibnamefont
  {Martin}}, \bibinfo {author} {\bibfnamefont {Y.-H.}\ \bibnamefont {Chu}},\
  and\ \bibinfo {author} {\bibfnamefont {R.}~\bibnamefont {Ramesh}},\
  }\bibfield  {title} {\bibinfo {title} {Advances in the growth and
  characterization of magnetic, ferroelectric, and multiferroic oxide thin
  films},\ }\href {https://doi.org/https://doi.org/10.1016/j.mser.2010.03.001}
  {\bibfield  {journal} {\bibinfo  {journal} {Materials Science and
  Engineering: R: Reports}\ }\textbf {\bibinfo {volume} {68}},\ \bibinfo
  {pages} {89 } (\bibinfo {year} {2010})}\BibitemShut {NoStop}%
\bibitem [{\citenamefont {Yamada}\ \emph {et~al.}(2013)\citenamefont {Yamada},
  \citenamefont {Garcia}, \citenamefont {Fusil}, \citenamefont {Boyn},
  \citenamefont {Marinova}, \citenamefont {Gloter}, \citenamefont {Xavier},
  \citenamefont {Grollier}, \citenamefont {Jacquet}, \citenamefont
  {Carr{\'e}t{\'e}ro}, \citenamefont {Deranlot}, \citenamefont {Bibes},\ and\
  \citenamefont {Barth{\'e}l{\'e}my}}]{Yamada2013}%
  \BibitemOpen
  \bibfield  {author} {\bibinfo {author} {\bibfnamefont {H.}~\bibnamefont
  {Yamada}}, \bibinfo {author} {\bibfnamefont {V.}~\bibnamefont {Garcia}},
  \bibinfo {author} {\bibfnamefont {S.}~\bibnamefont {Fusil}}, \bibinfo
  {author} {\bibfnamefont {S.}~\bibnamefont {Boyn}}, \bibinfo {author}
  {\bibfnamefont {M.}~\bibnamefont {Marinova}}, \bibinfo {author}
  {\bibfnamefont {A.}~\bibnamefont {Gloter}}, \bibinfo {author} {\bibfnamefont
  {S.}~\bibnamefont {Xavier}}, \bibinfo {author} {\bibfnamefont
  {J.}~\bibnamefont {Grollier}}, \bibinfo {author} {\bibfnamefont
  {E.}~\bibnamefont {Jacquet}}, \bibinfo {author} {\bibfnamefont
  {C.}~\bibnamefont {Carr{\'e}t{\'e}ro}}, \bibinfo {author} {\bibfnamefont
  {C.}~\bibnamefont {Deranlot}}, \bibinfo {author} {\bibfnamefont
  {M.}~\bibnamefont {Bibes}},\ and\ \bibinfo {author} {\bibfnamefont
  {A.}~\bibnamefont {Barth{\'e}l{\'e}my}},\ }\bibfield  {title} {\bibinfo
  {title} {Giant electroresistance of super-tetragonal {BiFeO${}_{3}$}-based
  ferroelectric tunnel junctions},\ }\href {https://doi.org/10.1021/nn401378t}
  {\bibfield  {journal} {\bibinfo  {journal} {ACS Nano}\ }\textbf {\bibinfo
  {volume} {7}},\ \bibinfo {pages} {5385} (\bibinfo {year} {2013})},\ \bibinfo
  {note} {pMID: 23647323},\ \Eprint
  {https://arxiv.org/abs/https://doi.org/10.1021/nn401378t}
  {https://doi.org/10.1021/nn401378t} \BibitemShut {NoStop}%
\bibitem [{\citenamefont {Qu}\ \emph {et~al.}(2012)\citenamefont {Qu},
  \citenamefont {Zhao}, \citenamefont {Yu}, \citenamefont {Zhao}, \citenamefont
  {Zhang},\ and\ \citenamefont {Yang}}]{Qu2012}%
  \BibitemOpen
  \bibfield  {author} {\bibinfo {author} {\bibfnamefont {T.~L.}\ \bibnamefont
  {Qu}}, \bibinfo {author} {\bibfnamefont {Y.~G.}\ \bibnamefont {Zhao}},
  \bibinfo {author} {\bibfnamefont {P.}~\bibnamefont {Yu}}, \bibinfo {author}
  {\bibfnamefont {H.~C.}\ \bibnamefont {Zhao}}, \bibinfo {author}
  {\bibfnamefont {S.}~\bibnamefont {Zhang}},\ and\ \bibinfo {author}
  {\bibfnamefont {L.~F.}\ \bibnamefont {Yang}},\ }\bibfield  {title} {\bibinfo
  {title} {Exchange bias effects in epitaxial
  {Fe${}_{3}$O${}_{4}$/BiFeO${}_{3}$} heterostructures},\ }\href
  {https://doi.org/10.1063/1.4729408} {\bibfield  {journal} {\bibinfo
  {journal} {Applied Physics Letters}\ }\textbf {\bibinfo {volume} {100}},\
  \bibinfo {pages} {242410} (\bibinfo {year} {2012})},\ \Eprint
  {https://arxiv.org/abs/https://doi.org/10.1063/1.4729408}
  {https://doi.org/10.1063/1.4729408} \BibitemShut {NoStop}%
\bibitem [{\citenamefont {Chakrabarti}\ \emph {et~al.}(2014)\citenamefont
  {Chakrabarti}, \citenamefont {Sarkar}, \citenamefont {Dev~Ashok},
  \citenamefont {Das}, \citenamefont {Sinha~Chaudhuri}, \citenamefont {Mitra},\
  and\ \citenamefont {De}}]{Chakrabarti2014}%
  \BibitemOpen
  \bibfield  {author} {\bibinfo {author} {\bibfnamefont {K.}~\bibnamefont
  {Chakrabarti}}, \bibinfo {author} {\bibfnamefont {B.}~\bibnamefont {Sarkar}},
  \bibinfo {author} {\bibfnamefont {V.}~\bibnamefont {Dev~Ashok}}, \bibinfo
  {author} {\bibfnamefont {K.}~\bibnamefont {Das}}, \bibinfo {author}
  {\bibfnamefont {S.}~\bibnamefont {Sinha~Chaudhuri}}, \bibinfo {author}
  {\bibfnamefont {A.}~\bibnamefont {Mitra}},\ and\ \bibinfo {author}
  {\bibfnamefont {S.~K.}\ \bibnamefont {De}},\ }\bibfield  {title} {\bibinfo
  {title} {Exchange bias effect in {BiFeO${}_{3}$-NiO} nanocomposite},\ }\href
  {https://doi.org/10.1063/1.4861140} {\bibfield  {journal} {\bibinfo
  {journal} {Journal of Applied Physics}\ }\textbf {\bibinfo {volume} {115}},\
  \bibinfo {pages} {013906} (\bibinfo {year} {2014})},\ \Eprint
  {https://arxiv.org/abs/https://doi.org/10.1063/1.4861140}
  {https://doi.org/10.1063/1.4861140} \BibitemShut {NoStop}%
\bibitem [{\citenamefont {Manipatruni}\ \emph {et~al.}(2019)\citenamefont
  {Manipatruni}, \citenamefont {Nikonov}, \citenamefont {Lin}, \citenamefont
  {Gosavi}, \citenamefont {Liu}, \citenamefont {Prasad}, \citenamefont {Huang},
  \citenamefont {Bonturim}, \citenamefont {Ramesh},\ and\ \citenamefont
  {Young}}]{manipatruni2019}%
  \BibitemOpen
  \bibfield  {author} {\bibinfo {author} {\bibfnamefont {S.}~\bibnamefont
  {Manipatruni}}, \bibinfo {author} {\bibfnamefont {D.~E.}\ \bibnamefont
  {Nikonov}}, \bibinfo {author} {\bibfnamefont {C.-C.}\ \bibnamefont {Lin}},
  \bibinfo {author} {\bibfnamefont {T.~A.}\ \bibnamefont {Gosavi}}, \bibinfo
  {author} {\bibfnamefont {H.}~\bibnamefont {Liu}}, \bibinfo {author}
  {\bibfnamefont {B.}~\bibnamefont {Prasad}}, \bibinfo {author} {\bibfnamefont
  {Y.-L.}\ \bibnamefont {Huang}}, \bibinfo {author} {\bibfnamefont
  {E.}~\bibnamefont {Bonturim}}, \bibinfo {author} {\bibfnamefont
  {R.}~\bibnamefont {Ramesh}},\ and\ \bibinfo {author} {\bibfnamefont {I.~A.}\
  \bibnamefont {Young}},\ }\bibfield  {title} {\bibinfo {title} {Scalable
  energy-efficient magnetoelectric spin--orbit logic},\ }\href@noop {}
  {\bibfield  {journal} {\bibinfo  {journal} {Nature}\ }\textbf {\bibinfo
  {volume} {565}},\ \bibinfo {pages} {35} (\bibinfo {year} {2019})}\BibitemShut
  {NoStop}%
\bibitem [{Note1()}]{Note1}%
  \BibitemOpen
  \bibinfo {note} {In this paper the crystallographic directions are described
  in the pseudo-cubic notation.}\BibitemShut {Stop}%
\bibitem [{\citenamefont {Ito}\ \emph {et~al.}(2011)\citenamefont {Ito},
  \citenamefont {Ushiyama}, \citenamefont {Yanagisawa}, \citenamefont {Kumai},\
  and\ \citenamefont {Tomioka}}]{Ito2011}%
  \BibitemOpen
  \bibfield  {author} {\bibinfo {author} {\bibfnamefont {T.}~\bibnamefont
  {Ito}}, \bibinfo {author} {\bibfnamefont {T.}~\bibnamefont {Ushiyama}},
  \bibinfo {author} {\bibfnamefont {Y.}~\bibnamefont {Yanagisawa}}, \bibinfo
  {author} {\bibfnamefont {R.}~\bibnamefont {Kumai}},\ and\ \bibinfo {author}
  {\bibfnamefont {Y.}~\bibnamefont {Tomioka}},\ }\bibfield  {title} {\bibinfo
  {title} {Growth of highly insulating bulk single crystals of multiferroic
  {BiFeO$_3$} and their inherent internal strains in the domain-switching
  process},\ }\href {https://doi.org/10.1021/cg201068m} {\bibfield  {journal}
  {\bibinfo  {journal} {Crystal Growth \& Design}\ }\textbf {\bibinfo {volume}
  {11}},\ \bibinfo {pages} {5139} (\bibinfo {year} {2011})},\ \Eprint
  {https://arxiv.org/abs/http://dx.doi.org/10.1021/cg201068m}
  {http://dx.doi.org/10.1021/cg201068m} \BibitemShut {NoStop}%
\bibitem [{\citenamefont {Matsuda}\ \emph {et~al.}(2012)\citenamefont
  {Matsuda}, \citenamefont {Fishman}, \citenamefont {Hong}, \citenamefont
  {Lee}, \citenamefont {Ushiyama}, \citenamefont {Yanagisawa}, \citenamefont
  {Tomioka},\ and\ \citenamefont {Ito}}]{Matsuda2012}%
  \BibitemOpen
  \bibfield  {author} {\bibinfo {author} {\bibfnamefont {M.}~\bibnamefont
  {Matsuda}}, \bibinfo {author} {\bibfnamefont {R.~S.}\ \bibnamefont
  {Fishman}}, \bibinfo {author} {\bibfnamefont {T.}~\bibnamefont {Hong}},
  \bibinfo {author} {\bibfnamefont {C.~H.}\ \bibnamefont {Lee}}, \bibinfo
  {author} {\bibfnamefont {T.}~\bibnamefont {Ushiyama}}, \bibinfo {author}
  {\bibfnamefont {Y.}~\bibnamefont {Yanagisawa}}, \bibinfo {author}
  {\bibfnamefont {Y.}~\bibnamefont {Tomioka}},\ and\ \bibinfo {author}
  {\bibfnamefont {T.}~\bibnamefont {Ito}},\ }\bibfield  {title} {\bibinfo
  {title} {Magnetic dispersion and anisotropy in multiferroic {BiFeO}$_3$},\
  }\href {https://doi.org/10.1103/PhysRevLett.109.067205} {\bibfield  {journal}
  {\bibinfo  {journal} {Phys. Rev. Lett.}\ }\textbf {\bibinfo {volume} {109}},\
  \bibinfo {pages} {067205} (\bibinfo {year} {2012})}\BibitemShut {NoStop}%
\bibitem [{\citenamefont {Jeong}\ \emph {et~al.}(2012)\citenamefont {Jeong},
  \citenamefont {Goremychkin}, \citenamefont {Guidi}, \citenamefont {Nakajima},
  \citenamefont {Jeon}, \citenamefont {Kim}, \citenamefont {Furukawa},
  \citenamefont {Kim}, \citenamefont {Lee}, \citenamefont {Kiryukhin},
  \citenamefont {Cheong},\ and\ \citenamefont {Park}}]{Jeong2012}%
  \BibitemOpen
  \bibfield  {author} {\bibinfo {author} {\bibfnamefont {J.}~\bibnamefont
  {Jeong}}, \bibinfo {author} {\bibfnamefont {E.~A.}\ \bibnamefont
  {Goremychkin}}, \bibinfo {author} {\bibfnamefont {T.}~\bibnamefont {Guidi}},
  \bibinfo {author} {\bibfnamefont {K.}~\bibnamefont {Nakajima}}, \bibinfo
  {author} {\bibfnamefont {G.~S.}\ \bibnamefont {Jeon}}, \bibinfo {author}
  {\bibfnamefont {S.-A.}\ \bibnamefont {Kim}}, \bibinfo {author} {\bibfnamefont
  {S.}~\bibnamefont {Furukawa}}, \bibinfo {author} {\bibfnamefont {Y.~B.}\
  \bibnamefont {Kim}}, \bibinfo {author} {\bibfnamefont {S.}~\bibnamefont
  {Lee}}, \bibinfo {author} {\bibfnamefont {V.}~\bibnamefont {Kiryukhin}},
  \bibinfo {author} {\bibfnamefont {S.-W.}\ \bibnamefont {Cheong}},\ and\
  \bibinfo {author} {\bibfnamefont {J.-G.}\ \bibnamefont {Park}},\ }\bibfield
  {title} {\bibinfo {title} {Spin wave measurements over the full brillouin
  zone of multiferroic {BiFeO$_3$}},\ }\href
  {https://doi.org/10.1103/PhysRevLett.108.077202} {\bibfield  {journal}
  {\bibinfo  {journal} {Phys. Rev. Lett.}\ }\textbf {\bibinfo {volume} {108}},\
  \bibinfo {pages} {077202} (\bibinfo {year} {2012})}\BibitemShut {NoStop}%
\bibitem [{\citenamefont {Cazayous}\ \emph {et~al.}(2008)\citenamefont
  {Cazayous}, \citenamefont {Gallais}, \citenamefont {Sacuto}, \citenamefont
  {de~Sousa}, \citenamefont {Lebeugle},\ and\ \citenamefont
  {Colson}}]{Cazayous2008}%
  \BibitemOpen
  \bibfield  {author} {\bibinfo {author} {\bibfnamefont {M.}~\bibnamefont
  {Cazayous}}, \bibinfo {author} {\bibfnamefont {Y.}~\bibnamefont {Gallais}},
  \bibinfo {author} {\bibfnamefont {A.}~\bibnamefont {Sacuto}}, \bibinfo
  {author} {\bibfnamefont {R.}~\bibnamefont {de~Sousa}}, \bibinfo {author}
  {\bibfnamefont {D.}~\bibnamefont {Lebeugle}},\ and\ \bibinfo {author}
  {\bibfnamefont {D.}~\bibnamefont {Colson}},\ }\bibfield  {title} {\bibinfo
  {title} {Possible observation of cycloidal electromagnons in {BiFeO$_3$}},\
  }\href {https://doi.org/10.1103/PhysRevLett.101.037601} {\bibfield  {journal}
  {\bibinfo  {journal} {Phys. Rev. Lett.}\ }\textbf {\bibinfo {volume} {101}},\
  \bibinfo {pages} {037601} (\bibinfo {year} {2008})}\BibitemShut {NoStop}%
\bibitem [{\citenamefont {Talbayev}\ \emph {et~al.}(2011)\citenamefont
  {Talbayev}, \citenamefont {Trugman}, \citenamefont {Lee}, \citenamefont {Yi},
  \citenamefont {Cheong},\ and\ \citenamefont {Taylor}}]{Talbayev2011PRBBiFeO}%
  \BibitemOpen
  \bibfield  {author} {\bibinfo {author} {\bibfnamefont {D.}~\bibnamefont
  {Talbayev}}, \bibinfo {author} {\bibfnamefont {S.~A.}\ \bibnamefont
  {Trugman}}, \bibinfo {author} {\bibfnamefont {S.}~\bibnamefont {Lee}},
  \bibinfo {author} {\bibfnamefont {H.~T.}\ \bibnamefont {Yi}}, \bibinfo
  {author} {\bibfnamefont {S.-W.}\ \bibnamefont {Cheong}},\ and\ \bibinfo
  {author} {\bibfnamefont {A.~J.}\ \bibnamefont {Taylor}},\ }\bibfield  {title}
  {\bibinfo {title} {Long-wavelength magnetic and magnetoelectric excitations
  in the ferroelectric antiferromagnet {BiFeO$_3$}},\ }\href
  {https://doi.org/10.1103/PhysRevB.83.094403} {\bibfield  {journal} {\bibinfo
  {journal} {Phys. Rev. B}\ }\textbf {\bibinfo {volume} {83}},\ \bibinfo
  {pages} {094403} (\bibinfo {year} {2011})}\BibitemShut {NoStop}%
\bibitem [{\citenamefont {Nagel}\ \emph {et~al.}(2013)\citenamefont {Nagel},
  \citenamefont {Fishman}, \citenamefont {Katuwal}, \citenamefont {Engelkamp},
  \citenamefont {Talbayev}, \citenamefont {Yi}, \citenamefont {Cheong},\ and\
  \citenamefont {R{\~o}{\~o}m}}]{Nagel2013}%
  \BibitemOpen
  \bibfield  {author} {\bibinfo {author} {\bibfnamefont {U.}~\bibnamefont
  {Nagel}}, \bibinfo {author} {\bibfnamefont {R.~S.}\ \bibnamefont {Fishman}},
  \bibinfo {author} {\bibfnamefont {T.}~\bibnamefont {Katuwal}}, \bibinfo
  {author} {\bibfnamefont {H.}~\bibnamefont {Engelkamp}}, \bibinfo {author}
  {\bibfnamefont {D.}~\bibnamefont {Talbayev}}, \bibinfo {author}
  {\bibfnamefont {H.~T.}\ \bibnamefont {Yi}}, \bibinfo {author} {\bibfnamefont
  {S.-W.}\ \bibnamefont {Cheong}},\ and\ \bibinfo {author} {\bibfnamefont
  {T.}~\bibnamefont {R{\~o}{\~o}m}},\ }\bibfield  {title} {\bibinfo {title}
  {Terahertz spectroscopy of spin waves in multiferroic {BiFeO$_3$} in high
  magnetic fields},\ }\href {https://doi.org/10.1103/PhysRevLett.110.257201}
  {\bibfield  {journal} {\bibinfo  {journal} {Phys. Rev. Lett.}\ }\textbf
  {\bibinfo {volume} {110}},\ \bibinfo {pages} {257201} (\bibinfo {year}
  {2013})}\BibitemShut {NoStop}%
\bibitem [{\citenamefont {R{\~o}{\~o}m}\ \emph {et~al.}(2020)\citenamefont
  {R{\~o}{\~o}m}, \citenamefont {Viirok}, \citenamefont {Peedu}, \citenamefont
  {Nagel}, \citenamefont {Farkas}, \citenamefont {Szaller}, \citenamefont
  {Kocsis}, \citenamefont {Bord\'acs}, \citenamefont {K\'ezsm\'arki},
  \citenamefont {Kamenskyi}, \citenamefont {Engelkamp}, \citenamefont {Ozerov},
  \citenamefont {Smirnov}, \citenamefont {Krzystek}, \citenamefont
  {Thirunavukkuarasu}, \citenamefont {Ozaki}, \citenamefont {Tomioka},
  \citenamefont {Ito}, \citenamefont {Datta},\ and\ \citenamefont
  {Fishman}}]{Room2020highfield}%
  \BibitemOpen
  \bibfield  {author} {\bibinfo {author} {\bibfnamefont {T.}~\bibnamefont
  {R{\~o}{\~o}m}}, \bibinfo {author} {\bibfnamefont {J.}~\bibnamefont
  {Viirok}}, \bibinfo {author} {\bibfnamefont {L.}~\bibnamefont {Peedu}},
  \bibinfo {author} {\bibfnamefont {U.}~\bibnamefont {Nagel}}, \bibinfo
  {author} {\bibfnamefont {D.~G.}\ \bibnamefont {Farkas}}, \bibinfo {author}
  {\bibfnamefont {D.}~\bibnamefont {Szaller}}, \bibinfo {author} {\bibfnamefont
  {V.}~\bibnamefont {Kocsis}}, \bibinfo {author} {\bibfnamefont
  {S.}~\bibnamefont {Bord\'acs}}, \bibinfo {author} {\bibfnamefont
  {I.}~\bibnamefont {K\'ezsm\'arki}}, \bibinfo {author} {\bibfnamefont {D.~L.}\
  \bibnamefont {Kamenskyi}}, \bibinfo {author} {\bibfnamefont {H.}~\bibnamefont
  {Engelkamp}}, \bibinfo {author} {\bibfnamefont {M.}~\bibnamefont {Ozerov}},
  \bibinfo {author} {\bibfnamefont {D.}~\bibnamefont {Smirnov}}, \bibinfo
  {author} {\bibfnamefont {J.}~\bibnamefont {Krzystek}}, \bibinfo {author}
  {\bibfnamefont {K.}~\bibnamefont {Thirunavukkuarasu}}, \bibinfo {author}
  {\bibfnamefont {Y.}~\bibnamefont {Ozaki}}, \bibinfo {author} {\bibfnamefont
  {Y.}~\bibnamefont {Tomioka}}, \bibinfo {author} {\bibfnamefont
  {T.}~\bibnamefont {Ito}}, \bibinfo {author} {\bibfnamefont {T.}~\bibnamefont
  {Datta}},\ and\ \bibinfo {author} {\bibfnamefont {R.~S.}\ \bibnamefont
  {Fishman}},\ }\bibfield  {title} {\bibinfo {title} {Magnetoelastic distortion
  of multiferroic {BiFeO$_3$} in the canted antiferromagnetic state},\ }\href
  {https://doi.org/10.1103/PhysRevB.102.214410} {\bibfield  {journal} {\bibinfo
   {journal} {Phys. Rev. B}\ }\textbf {\bibinfo {volume} {102}},\ \bibinfo
  {pages} {214410} (\bibinfo {year} {2020})}\BibitemShut {NoStop}%
\bibitem [{\citenamefont {Fishman}\ \emph {et~al.}(2013)\citenamefont
  {Fishman}, \citenamefont {Haraldsen}, \citenamefont {Furukawa},\ and\
  \citenamefont {Miyahara}}]{Fishman2013PRB}%
  \BibitemOpen
  \bibfield  {author} {\bibinfo {author} {\bibfnamefont {R.~S.}\ \bibnamefont
  {Fishman}}, \bibinfo {author} {\bibfnamefont {J.~T.}\ \bibnamefont
  {Haraldsen}}, \bibinfo {author} {\bibfnamefont {N.}~\bibnamefont
  {Furukawa}},\ and\ \bibinfo {author} {\bibfnamefont {S.}~\bibnamefont
  {Miyahara}},\ }\bibfield  {title} {\bibinfo {title} {Spin state and
  spectroscopic modes of multiferroic {BiFeO$_3$}},\ }\href
  {https://doi.org/10.1103/PhysRevB.87.134416} {\bibfield  {journal} {\bibinfo
  {journal} {Phys. Rev. B}\ }\textbf {\bibinfo {volume} {87}},\ \bibinfo
  {pages} {134416} (\bibinfo {year} {2013})}\BibitemShut {NoStop}%
\bibitem [{\citenamefont {Fishman}(2013)}]{Fishman2013}%
  \BibitemOpen
  \bibfield  {author} {\bibinfo {author} {\bibfnamefont {R.~S.}\ \bibnamefont
  {Fishman}},\ }\bibfield  {title} {\bibinfo {title} {Field dependence of the
  spin state and spectroscopic modes of multiferroic {BiFeO$_3$}},\ }\href
  {https://doi.org/10.1103/PhysRevB.87.224419} {\bibfield  {journal} {\bibinfo
  {journal} {Phys. Rev. B}\ }\textbf {\bibinfo {volume} {87}},\ \bibinfo
  {pages} {224419} (\bibinfo {year} {2013})}\BibitemShut {NoStop}%
\bibitem [{\citenamefont {Popov}\ \emph {et~al.}(1993)\citenamefont {Popov},
  \citenamefont {Zvezdin}, \citenamefont {Vorob'ev}, \citenamefont
  {Kadomtseva}, \citenamefont {Murashev},\ and\ \citenamefont
  {Rakov}}]{Popov1993}%
  \BibitemOpen
  \bibfield  {author} {\bibinfo {author} {\bibfnamefont {Y.~F.}\ \bibnamefont
  {Popov}}, \bibinfo {author} {\bibfnamefont {A.~K.}\ \bibnamefont {Zvezdin}},
  \bibinfo {author} {\bibfnamefont {G.~P.}\ \bibnamefont {Vorob'ev}}, \bibinfo
  {author} {\bibfnamefont {A.~M.}\ \bibnamefont {Kadomtseva}}, \bibinfo
  {author} {\bibfnamefont {V.~A.}\ \bibnamefont {Murashev}},\ and\ \bibinfo
  {author} {\bibfnamefont {D.~N.}\ \bibnamefont {Rakov}},\ }\bibfield  {title}
  {\bibinfo {title} {Linear magnetoelectric effect and phase transitions in
  bismuth ferrite, ${\text{bifeo}}_{3}$},\ }\href@noop {} {\bibfield  {journal}
  {\bibinfo  {journal} {JETP Lett.}\ }\textbf {\bibinfo {volume} {57}},\
  \bibinfo {pages} {65} (\bibinfo {year} {1993})}\BibitemShut {NoStop}%
\bibitem [{\citenamefont {Popov}\ \emph {et~al.}(2001)\citenamefont {Popov},
  \citenamefont {Kadomtseva}, \citenamefont {Krotov}, \citenamefont {Belov},
  \citenamefont {Vorob'ev}, \citenamefont {Makhov},\ and\ \citenamefont
  {Zvezdin}}]{Popov2001}%
  \BibitemOpen
  \bibfield  {author} {\bibinfo {author} {\bibfnamefont {Y.~F.}\ \bibnamefont
  {Popov}}, \bibinfo {author} {\bibfnamefont {A.~M.}\ \bibnamefont
  {Kadomtseva}}, \bibinfo {author} {\bibfnamefont {S.~S.}\ \bibnamefont
  {Krotov}}, \bibinfo {author} {\bibfnamefont {D.~V.}\ \bibnamefont {Belov}},
  \bibinfo {author} {\bibfnamefont {G.~P.}\ \bibnamefont {Vorob'ev}}, \bibinfo
  {author} {\bibfnamefont {P.~N.}\ \bibnamefont {Makhov}},\ and\ \bibinfo
  {author} {\bibfnamefont {A.~K.}\ \bibnamefont {Zvezdin}},\ }\bibfield
  {title} {\bibinfo {title} {Features of the magnetoelectric properties of
  ${\text{bifeo}}_{3}$ in high magnetic fields},\ }\href@noop {} {\bibfield
  {journal} {\bibinfo  {journal} {Low Temp. Phys.}\ }\textbf {\bibinfo {volume}
  {27}},\ \bibinfo {pages} {478} (\bibinfo {year} {2001})}\BibitemShut
  {NoStop}%
\bibitem [{\citenamefont {Tokunaga}\ \emph {et~al.}(2015)\citenamefont
  {Tokunaga}, \citenamefont {Akaki}, \citenamefont {Ito}, \citenamefont
  {Miyahara}, \citenamefont {Miyake}, \citenamefont {Kuwahara},\ and\
  \citenamefont {Furukawa}}]{Tokunaga2015}%
  \BibitemOpen
  \bibfield  {author} {\bibinfo {author} {\bibfnamefont {M.}~\bibnamefont
  {Tokunaga}}, \bibinfo {author} {\bibfnamefont {M.}~\bibnamefont {Akaki}},
  \bibinfo {author} {\bibfnamefont {T.}~\bibnamefont {Ito}}, \bibinfo {author}
  {\bibfnamefont {S.}~\bibnamefont {Miyahara}}, \bibinfo {author}
  {\bibfnamefont {A.}~\bibnamefont {Miyake}}, \bibinfo {author} {\bibfnamefont
  {H.}~\bibnamefont {Kuwahara}},\ and\ \bibinfo {author} {\bibfnamefont
  {N.}~\bibnamefont {Furukawa}},\ }\bibfield  {title} {\bibinfo {title}
  {Magnetic control of transverse electric polarization in {BiFeO$_3$}},\
  }\href@noop {} {\bibfield  {journal} {\bibinfo  {journal} {Nature
  Communications}\ }\textbf {\bibinfo {volume} {6}},\ \bibinfo {pages} {5878}
  (\bibinfo {year} {2015})}\BibitemShut {NoStop}%
\bibitem [{\citenamefont {Bord\'acs}\ \emph {et~al.}(2018)\citenamefont
  {Bord\'acs}, \citenamefont {Farkas}, \citenamefont {White}, \citenamefont
  {Cubitt}, \citenamefont {DeBeer-Schmitt}, \citenamefont {Ito},\ and\
  \citenamefont {K\'ezsm\'arki}}]{Bordacs2018}%
  \BibitemOpen
  \bibfield  {author} {\bibinfo {author} {\bibfnamefont {S.}~\bibnamefont
  {Bord\'acs}}, \bibinfo {author} {\bibfnamefont {D.~G.}\ \bibnamefont
  {Farkas}}, \bibinfo {author} {\bibfnamefont {J.~S.}\ \bibnamefont {White}},
  \bibinfo {author} {\bibfnamefont {R.}~\bibnamefont {Cubitt}}, \bibinfo
  {author} {\bibfnamefont {L.}~\bibnamefont {DeBeer-Schmitt}}, \bibinfo
  {author} {\bibfnamefont {T.}~\bibnamefont {Ito}},\ and\ \bibinfo {author}
  {\bibfnamefont {I.}~\bibnamefont {K\'ezsm\'arki}},\ }\bibfield  {title}
  {\bibinfo {title} {Magnetic field control of cycloidal domains and electric
  polarization in multiferroic {BiFeO}$_{3}$},\ }\href
  {https://doi.org/10.1103/PhysRevLett.120.147203} {\bibfield  {journal}
  {\bibinfo  {journal} {Phys. Rev. Lett.}\ }\textbf {\bibinfo {volume} {120}},\
  \bibinfo {pages} {147203} (\bibinfo {year} {2018})}\BibitemShut {NoStop}%
\bibitem [{\citenamefont {Kawachi}\ \emph {et~al.}(2019)\citenamefont
  {Kawachi}, \citenamefont {Miyahara}, \citenamefont {Ito}, \citenamefont
  {Miyake}, \citenamefont {Furukawa}, \citenamefont {Yamaura},\ and\
  \citenamefont {Tokunaga}}]{kawachi2019direct}%
  \BibitemOpen
  \bibfield  {author} {\bibinfo {author} {\bibfnamefont {S.}~\bibnamefont
  {Kawachi}}, \bibinfo {author} {\bibfnamefont {S.}~\bibnamefont {Miyahara}},
  \bibinfo {author} {\bibfnamefont {T.}~\bibnamefont {Ito}}, \bibinfo {author}
  {\bibfnamefont {A.}~\bibnamefont {Miyake}}, \bibinfo {author} {\bibfnamefont
  {N.}~\bibnamefont {Furukawa}}, \bibinfo {author} {\bibfnamefont {J.-i.}\
  \bibnamefont {Yamaura}},\ and\ \bibinfo {author} {\bibfnamefont
  {M.}~\bibnamefont {Tokunaga}},\ }\bibfield  {title} {\bibinfo {title} {Direct
  coupling of ferromagnetic moment and ferroelectric polarization in
  {BiFeO$_3$}},\ }\href@noop {} {\bibfield  {journal} {\bibinfo  {journal}
  {Physical Review B}\ }\textbf {\bibinfo {volume} {100}},\ \bibinfo {pages}
  {140412(R)} (\bibinfo {year} {2019})}\BibitemShut {NoStop}%
\bibitem [{\citenamefont {K\'ezsm\'arki}\ \emph {et~al.}(2015)\citenamefont
  {K\'ezsm\'arki}, \citenamefont {Nagel}, \citenamefont {Bord\'acs},
  \citenamefont {Fishman}, \citenamefont {Lee}, \citenamefont {Yi},
  \citenamefont {Cheong},\ and\ \citenamefont {R{\~o\~o}m}}]{Kezsmarki2015}%
  \BibitemOpen
  \bibfield  {author} {\bibinfo {author} {\bibfnamefont {I.}~\bibnamefont
  {K\'ezsm\'arki}}, \bibinfo {author} {\bibfnamefont {U.}~\bibnamefont
  {Nagel}}, \bibinfo {author} {\bibfnamefont {S.}~\bibnamefont {Bord\'acs}},
  \bibinfo {author} {\bibfnamefont {R.~S.}\ \bibnamefont {Fishman}}, \bibinfo
  {author} {\bibfnamefont {J.~H.}\ \bibnamefont {Lee}}, \bibinfo {author}
  {\bibfnamefont {H.~T.}\ \bibnamefont {Yi}}, \bibinfo {author} {\bibfnamefont
  {S.-W.}\ \bibnamefont {Cheong}},\ and\ \bibinfo {author} {\bibfnamefont
  {T.}~\bibnamefont {R{\~o\~o}m}},\ }\bibfield  {title} {\bibinfo {title}
  {Optical diode effect at spin-wave excitations of the room-temperature
  multiferroic {${\mathrm{BiFeO}}_{3}$}},\ }\href
  {https://doi.org/10.1103/PhysRevLett.115.127203} {\bibfield  {journal}
  {\bibinfo  {journal} {Phys. Rev. Lett.}\ }\textbf {\bibinfo {volume} {115}},\
  \bibinfo {pages} {127203} (\bibinfo {year} {2015})}\BibitemShut {NoStop}%
\bibitem [{\citenamefont {K{\'e}zsm{\'a}rki}\ \emph {et~al.}(2011)\citenamefont
  {K{\'e}zsm{\'a}rki}, \citenamefont {Kida}, \citenamefont {Murakawa},
  \citenamefont {Bord{\'a}cs}, \citenamefont {Onose},\ and\ \citenamefont
  {Tokura}}]{Kezsmarki2011}%
  \BibitemOpen
  \bibfield  {author} {\bibinfo {author} {\bibfnamefont {I.}~\bibnamefont
  {K{\'e}zsm{\'a}rki}}, \bibinfo {author} {\bibfnamefont {N.}~\bibnamefont
  {Kida}}, \bibinfo {author} {\bibfnamefont {H.}~\bibnamefont {Murakawa}},
  \bibinfo {author} {\bibfnamefont {S.}~\bibnamefont {Bord{\'a}cs}}, \bibinfo
  {author} {\bibfnamefont {Y.}~\bibnamefont {Onose}},\ and\ \bibinfo {author}
  {\bibfnamefont {Y.}~\bibnamefont {Tokura}},\ }\bibfield  {title} {\bibinfo
  {title} {Enhanced directional dichroism of terahertz light in resonance with
  magnetic excitations of the multiferroic {Ba$_2$CoGe$_2$O$_7$} oxide
  compound},\ }\href {https://doi.org/10.1103/PhysRevLett.106.057403}
  {\bibfield  {journal} {\bibinfo  {journal} {Phys. Rev. Lett.}\ }\textbf
  {\bibinfo {volume} {106}},\ \bibinfo {pages} {057403} (\bibinfo {year}
  {2011})}\BibitemShut {NoStop}%
\bibitem [{\citenamefont {Bord{\'a}cs}\ \emph {et~al.}(2012)\citenamefont
  {Bord{\'a}cs}, \citenamefont {K{\'e}zsm{\'a}rki}, \citenamefont {Szaller},
  \citenamefont {Demk{\'o}}, \citenamefont {Kida}, \citenamefont {Murakawa},
  \citenamefont {Onose}, \citenamefont {Shimano}, \citenamefont {R{\~o}{\~o}m},
  \citenamefont {Nagel}, \citenamefont {Miyahara}, \citenamefont {Furukawa},\
  and\ \citenamefont {Tokura}}]{Bordacs2012}%
  \BibitemOpen
  \bibfield  {author} {\bibinfo {author} {\bibfnamefont {S.}~\bibnamefont
  {Bord{\'a}cs}}, \bibinfo {author} {\bibfnamefont {I.}~\bibnamefont
  {K{\'e}zsm{\'a}rki}}, \bibinfo {author} {\bibfnamefont {D.}~\bibnamefont
  {Szaller}}, \bibinfo {author} {\bibfnamefont {L.}~\bibnamefont {Demk{\'o}}},
  \bibinfo {author} {\bibfnamefont {N.}~\bibnamefont {Kida}}, \bibinfo {author}
  {\bibfnamefont {H.}~\bibnamefont {Murakawa}}, \bibinfo {author}
  {\bibfnamefont {Y.}~\bibnamefont {Onose}}, \bibinfo {author} {\bibfnamefont
  {R.}~\bibnamefont {Shimano}}, \bibinfo {author} {\bibfnamefont
  {T.}~\bibnamefont {R{\~o}{\~o}m}}, \bibinfo {author} {\bibfnamefont
  {U.}~\bibnamefont {Nagel}}, \bibinfo {author} {\bibfnamefont
  {S.}~\bibnamefont {Miyahara}}, \bibinfo {author} {\bibfnamefont
  {N.}~\bibnamefont {Furukawa}},\ and\ \bibinfo {author} {\bibfnamefont
  {Y.}~\bibnamefont {Tokura}},\ }\bibfield  {title} {\bibinfo {title}
  {Chirality of matter shows up via spin excitations},\ }\href
  {https://doi.org/doi:10.1038/nphys2387} {\bibfield  {journal} {\bibinfo
  {journal} {Nature Physics}\ }\textbf {\bibinfo {volume} {8}},\ \bibinfo
  {pages} {734} (\bibinfo {year} {2012})}\BibitemShut {NoStop}%
\bibitem [{\citenamefont {Takahashi}\ \emph {et~al.}(2012)\citenamefont
  {Takahashi}, \citenamefont {Shimano}, \citenamefont {Kaneko}, \citenamefont
  {Murakawa},\ and\ \citenamefont {Tokura}}]{takahashi2012}%
  \BibitemOpen
  \bibfield  {author} {\bibinfo {author} {\bibfnamefont {Y.}~\bibnamefont
  {Takahashi}}, \bibinfo {author} {\bibfnamefont {R.}~\bibnamefont {Shimano}},
  \bibinfo {author} {\bibfnamefont {Y.}~\bibnamefont {Kaneko}}, \bibinfo
  {author} {\bibfnamefont {H.}~\bibnamefont {Murakawa}},\ and\ \bibinfo
  {author} {\bibfnamefont {Y.}~\bibnamefont {Tokura}},\ }\bibfield  {title}
  {\bibinfo {title} {Magnetoelectric resonance with electromagnons in a
  perovskite helimagnet},\ }\href@noop {} {\bibfield  {journal} {\bibinfo
  {journal} {Nature Physics}\ }\textbf {\bibinfo {volume} {8}},\ \bibinfo
  {pages} {121} (\bibinfo {year} {2012})}\BibitemShut {NoStop}%
\bibitem [{\citenamefont {Yu}\ \emph {et~al.}(2018)\citenamefont {Yu},
  \citenamefont {Gao}, \citenamefont {Kim}, \citenamefont {Cheong},
  \citenamefont {Man}, \citenamefont {Mad\'eo}, \citenamefont {Dani},\ and\
  \citenamefont {Talbayev}}]{yu2018}%
  \BibitemOpen
  \bibfield  {author} {\bibinfo {author} {\bibfnamefont {S.}~\bibnamefont
  {Yu}}, \bibinfo {author} {\bibfnamefont {B.}~\bibnamefont {Gao}}, \bibinfo
  {author} {\bibfnamefont {J.~W.}\ \bibnamefont {Kim}}, \bibinfo {author}
  {\bibfnamefont {S.-W.}\ \bibnamefont {Cheong}}, \bibinfo {author}
  {\bibfnamefont {M.~K.~L.}\ \bibnamefont {Man}}, \bibinfo {author}
  {\bibfnamefont {J.}~\bibnamefont {Mad\'eo}}, \bibinfo {author} {\bibfnamefont
  {K.~M.}\ \bibnamefont {Dani}},\ and\ \bibinfo {author} {\bibfnamefont
  {D.}~\bibnamefont {Talbayev}},\ }\bibfield  {title} {\bibinfo {title}
  {High-temperature terahertz optical diode effect without magnetic order in
  polar {FeZnMo$_3$O$_8$}},\ }\href
  {https://doi.org/10.1103/PhysRevLett.120.037601} {\bibfield  {journal}
  {\bibinfo  {journal} {Phys. Rev. Lett.}\ }\textbf {\bibinfo {volume} {120}},\
  \bibinfo {pages} {037601} (\bibinfo {year} {2018})}\BibitemShut {NoStop}%
\bibitem [{\citenamefont {Viirok}\ \emph {et~al.}(2019)\citenamefont {Viirok},
  \citenamefont {Nagel}, \citenamefont {R{\~o}{\~o}m}, \citenamefont {Farkas},
  \citenamefont {Balla}, \citenamefont {Szaller}, \citenamefont {Kocsis},
  \citenamefont {Tokunaga}, \citenamefont {Taguchi}, \citenamefont {Tokura}, \citenamefont {Bern\'ath}, \citenamefont {Kamenskyi}, \citenamefont {K\'ezsm\'arki}, \citenamefont {Bord\'acs}, \citenamefont {Penc}}]{viirok2019directional}%
  \BibitemOpen
  \bibfield  {author} {\bibinfo {author} {\bibfnamefont {J.}~\bibnamefont
  {Viirok}}, \bibinfo {author} {\bibfnamefont {U.}~\bibnamefont {Nagel}},
  \bibinfo {author} {\bibfnamefont {T.}~\bibnamefont {R{\~o}{\~o}m}}, \bibinfo
  {author} {\bibfnamefont {D.G.}~\bibnamefont {Farkas}}, \bibinfo {author}
  {\bibfnamefont {P.}~\bibnamefont {Balla}}, \bibinfo {author} {\bibfnamefont
  {D.}~\bibnamefont {Szaller}}, \bibinfo {author} {\bibfnamefont
  {V.}~\bibnamefont {Kocsis}}, \bibinfo {author} {\bibfnamefont
  {Y.}~\bibnamefont {Tokunaga}}, \bibinfo {author} {\bibfnamefont
  {Y.}~\bibnamefont {Taguchi}}, \bibinfo {author} {\bibfnamefont
  {Y.}~\bibnamefont {Tokura}}, \bibinfo {author} {\bibfnamefont
  {B.}~\bibnamefont {Bern\'ath}}, \bibinfo {author} {\bibfnamefont
  {D.L.}~\bibnamefont {Kamenskyi}}, \bibinfo {author} {\bibfnamefont
  {I.}~\bibnamefont {K\'ezsm\'arki}}, \bibinfo {author} {\bibfnamefont
  {S.}~\bibnamefont {Bord\'acs}}, \bibinfo {author} {\bibfnamefont
  {K.}~\bibnamefont {Penc}},}\bibfield  {title} {\bibinfo
  {title} {Directional dichroism in the paramagnetic state of multiferroics: A
  case study of infrared light absorption in sr$_2$cosi$_2$o$_7$ at high
  temperatures},\ }\href@noop {} {\bibfield  {journal} {\bibinfo  {journal}
  {Physical Review B}\ }\textbf {\bibinfo {volume} {99}},\ \bibinfo {pages}
  {014410} (\bibinfo {year} {2019})}\BibitemShut {NoStop}%
\bibitem [{\citenamefont {K{\'e}zsm{\'a}rki}\ \emph {et~al.}(2014)\citenamefont
  {K{\'e}zsm{\'a}rki}, \citenamefont {Szaller}, \citenamefont {Bord{\'a}cs},
  \citenamefont {Kocsis}, \citenamefont {Tokunaga}, \citenamefont {Taguchi},
  \citenamefont {Murakawa}, \citenamefont {Tokura}, \citenamefont {Engelkamp},
  \citenamefont {R{\~o}{\~o}m},\ and\ \citenamefont {Nagel}}]{Kezsmarki2014}%
  \BibitemOpen
  \bibfield  {author} {\bibinfo {author} {\bibfnamefont {I.}~\bibnamefont
  {K{\'e}zsm{\'a}rki}}, \bibinfo {author} {\bibfnamefont {D.}~\bibnamefont
  {Szaller}}, \bibinfo {author} {\bibfnamefont {S.}~\bibnamefont
  {Bord{\'a}cs}}, \bibinfo {author} {\bibfnamefont {V.}~\bibnamefont {Kocsis}},
  \bibinfo {author} {\bibfnamefont {Y.}~\bibnamefont {Tokunaga}}, \bibinfo
  {author} {\bibfnamefont {Y.}~\bibnamefont {Taguchi}}, \bibinfo {author}
  {\bibfnamefont {H.}~\bibnamefont {Murakawa}}, \bibinfo {author}
  {\bibfnamefont {Y.}~\bibnamefont {Tokura}}, \bibinfo {author} {\bibfnamefont
  {H.}~\bibnamefont {Engelkamp}}, \bibinfo {author} {\bibfnamefont
  {T.}~\bibnamefont {R{\~o}{\~o}m}},\ and\ \bibinfo {author} {\bibfnamefont
  {U.}~\bibnamefont {Nagel}},\ }\bibfield  {title} {\bibinfo {title} {One-way
  transparency of four-coloured spin-wave excitations in multiferroic
  materials},\ }\href {https://doi.org/10.1038/ncomms4203} {\bibfield
  {journal} {\bibinfo  {journal} {Nature Comm.}\ }\textbf {\bibinfo {volume}
  {5}},\ \bibinfo {pages} {3203} (\bibinfo {year} {2014})}\BibitemShut
  {NoStop}%
\bibitem [{\citenamefont {Peedu}\ \emph {et~al.}(2019)\citenamefont {Peedu},
  \citenamefont {Kocsis}, \citenamefont {Szaller}, \citenamefont {Viirok},
  \citenamefont {Nagel}, \citenamefont {R{\~o}{\~o}m}, \citenamefont {Farkas},
  \citenamefont {Bord\'acs}, \citenamefont {Kamenskyi}, \citenamefont
  {Zeitler}, \citenamefont {Tokunaga}, \citenamefont {Taguchi}, \citenamefont
  {Tokura},\ and\ \citenamefont {K\'ezsm\'arki}}]{Peedu2019LiNiPO}%
  \BibitemOpen
  \bibfield  {author} {\bibinfo {author} {\bibfnamefont {L.}~\bibnamefont
  {Peedu}}, \bibinfo {author} {\bibfnamefont {V.}~\bibnamefont {Kocsis}},
  \bibinfo {author} {\bibfnamefont {D.}~\bibnamefont {Szaller}}, \bibinfo
  {author} {\bibfnamefont {J.}~\bibnamefont {Viirok}}, \bibinfo {author}
  {\bibfnamefont {U.}~\bibnamefont {Nagel}}, \bibinfo {author} {\bibfnamefont
  {T.}~\bibnamefont {R{\~o}{\~o}m}}, \bibinfo {author} {\bibfnamefont {D.~G.}\
  \bibnamefont {Farkas}}, \bibinfo {author} {\bibfnamefont {S.}~\bibnamefont
  {Bord\'acs}}, \bibinfo {author} {\bibfnamefont {D.~L.}\ \bibnamefont
  {Kamenskyi}}, \bibinfo {author} {\bibfnamefont {U.}~\bibnamefont {Zeitler}},
  \bibinfo {author} {\bibfnamefont {Y.}~\bibnamefont {Tokunaga}}, \bibinfo
  {author} {\bibfnamefont {Y.}~\bibnamefont {Taguchi}}, \bibinfo {author}
  {\bibfnamefont {Y.}~\bibnamefont {Tokura}},\ and\ \bibinfo {author}
  {\bibfnamefont {I.}~\bibnamefont {K\'ezsm\'arki}},\ }\bibfield  {title}
  {\bibinfo {title} {Spin excitations of magnetoelectric
  $\mathrm{LiNiPO}{}_{4}$ in multiple magnetic phases},\ }\href
  {https://doi.org/10.1103/PhysRevB.100.024406} {\bibfield  {journal} {\bibinfo
   {journal} {Phys. Rev. B}\ }\textbf {\bibinfo {volume} {100}},\ \bibinfo
  {pages} {024406} (\bibinfo {year} {2019})}\BibitemShut {NoStop}%
\bibitem [{\citenamefont {Szaller}\ \emph {et~al.}(2017)\citenamefont
  {Szaller}, \citenamefont {Kocsis}, \citenamefont {Bord\'acs}, \citenamefont
  {Feh\'er}, \citenamefont {R{\~o}{\~o}m}, \citenamefont {Nagel}, \citenamefont
  {Engelkamp}, \citenamefont {Ohgushi},\ and\ \citenamefont
  {K\'ezsm\'arki}}]{Szaller2017TbFeBO}%
  \BibitemOpen
  \bibfield  {author} {\bibinfo {author} {\bibfnamefont {D.}~\bibnamefont
  {Szaller}}, \bibinfo {author} {\bibfnamefont {V.}~\bibnamefont {Kocsis}},
  \bibinfo {author} {\bibfnamefont {S.}~\bibnamefont {Bord\'acs}}, \bibinfo
  {author} {\bibfnamefont {T.}~\bibnamefont {Feh\'er}}, \bibinfo {author}
  {\bibfnamefont {T.}~\bibnamefont {R{\~o}{\~o}m}}, \bibinfo {author}
  {\bibfnamefont {U.}~\bibnamefont {Nagel}}, \bibinfo {author} {\bibfnamefont
  {H.}~\bibnamefont {Engelkamp}}, \bibinfo {author} {\bibfnamefont
  {K.}~\bibnamefont {Ohgushi}},\ and\ \bibinfo {author} {\bibfnamefont
  {I.}~\bibnamefont {K\'ezsm\'arki}},\ }\bibfield  {title} {\bibinfo {title}
  {Magnetic resonances of multiferroic tbfe$_3$(bo$_3$)$_4$},\ }\href
  {https://doi.org/10.1103/PhysRevB.95.024427} {\bibfield  {journal} {\bibinfo
  {journal} {Phys. Rev. B}\ }\textbf {\bibinfo {volume} {95}},\ \bibinfo
  {pages} {024427} (\bibinfo {year} {2017})}\BibitemShut {NoStop}%
\bibitem [{\citenamefont {Szaller}\ \emph {et~al.}(2020)\citenamefont
  {Szaller}, \citenamefont {Sz\'asz}, \citenamefont {Bord\'acs}, \citenamefont
  {Viirok}, \citenamefont {R{\~o}{\~o}m}, \citenamefont {Nagel}, \citenamefont
  {Shuvaev}, \citenamefont {Weymann}, \citenamefont {Pimenov}, \citenamefont
  {Tsirlin}, \citenamefont {Jesche}, \citenamefont {Prodan}, \citenamefont
  {Tsurkan},\ and\ \citenamefont {K\'ezsm\'arki}}]{Szaller2020MnMoO}%
  \BibitemOpen
  \bibfield  {author} {\bibinfo {author} {\bibfnamefont {D.}~\bibnamefont
  {Szaller}}, \bibinfo {author} {\bibfnamefont {K.}~\bibnamefont {Sz\'asz}},
  \bibinfo {author} {\bibfnamefont {S.}~\bibnamefont {Bord\'acs}}, \bibinfo
  {author} {\bibfnamefont {J.}~\bibnamefont {Viirok}}, \bibinfo {author}
  {\bibfnamefont {T.}~\bibnamefont {R{\~o}{\~o}m}}, \bibinfo {author}
  {\bibfnamefont {U.}~\bibnamefont {Nagel}}, \bibinfo {author} {\bibfnamefont
  {A.}~\bibnamefont {Shuvaev}}, \bibinfo {author} {\bibfnamefont
  {L.}~\bibnamefont {Weymann}}, \bibinfo {author} {\bibfnamefont
  {A.}~\bibnamefont {Pimenov}}, \bibinfo {author} {\bibfnamefont {A.~A.}\
  \bibnamefont {Tsirlin}}, \bibinfo {author} {\bibfnamefont {A.}~\bibnamefont
  {Jesche}}, \bibinfo {author} {\bibfnamefont {L.}~\bibnamefont {Prodan}},
  \bibinfo {author} {\bibfnamefont {V.}~\bibnamefont {Tsurkan}},\ and\ \bibinfo
  {author} {\bibfnamefont {I.}~\bibnamefont {K\'ezsm\'arki}},\ }\bibfield
  {title} {\bibinfo {title} {Magnetic anisotropy and exchange paths for
  octahedrally and tetrahedrally coordinated ${\mathrm{mn}}^{2+}$ ions in the
  honeycomb multiferroic mn$_2$mo$_3$o$_8$},\ }\href
  {https://doi.org/10.1103/PhysRevB.102.144410} {\bibfield  {journal} {\bibinfo
   {journal} {Phys. Rev. B}\ }\textbf {\bibinfo {volume} {102}},\ \bibinfo
  {pages} {144410} (\bibinfo {year} {2020})}\BibitemShut {NoStop}%
\bibitem [{\citenamefont {Tokura}(2007)}]{TOKURA20071145}%
  \BibitemOpen
  \bibfield  {author} {\bibinfo {author} {\bibfnamefont {Y.}~\bibnamefont
  {Tokura}},\ }\bibfield  {title} {\bibinfo {title} {Multiferroics—toward
  strong coupling between magnetization and polarization in a solid},\ }\href
  {https://doi.org/https://doi.org/10.1016/j.jmmm.2006.11.198} {\bibfield
  {journal} {\bibinfo  {journal} {Journal of Magnetism and Magnetic Materials}\
  }\textbf {\bibinfo {volume} {310}},\ \bibinfo {pages} {1145} (\bibinfo {year}
  {2007})},\ \bibinfo {note} {proceedings of the 17th International Conference
  on Magnetism}\BibitemShut {NoStop}%
\bibitem [{\citenamefont {Szaller}\ \emph {et~al.}(2013)\citenamefont
  {Szaller}, \citenamefont {Bord\'acs},\ and\ \citenamefont
  {K\'ezsm\'arki}}]{Szaller2013}%
  \BibitemOpen
  \bibfield  {author} {\bibinfo {author} {\bibfnamefont {D.}~\bibnamefont
  {Szaller}}, \bibinfo {author} {\bibfnamefont {S.}~\bibnamefont {Bord\'acs}},\
  and\ \bibinfo {author} {\bibfnamefont {I.}~\bibnamefont {K\'ezsm\'arki}},\
  }\bibfield  {title} {\bibinfo {title} {Symmetry conditions for nonreciprocal
  light propagation in magnetic crystals},\ }\href
  {https://doi.org/10.1103/PhysRevB.87.014421} {\bibfield  {journal} {\bibinfo
  {journal} {Phys. Rev. B}\ }\textbf {\bibinfo {volume} {87}},\ \bibinfo
  {pages} {014421} (\bibinfo {year} {2013})}\BibitemShut {NoStop}%
\bibitem [{\citenamefont {Fishman}\ \emph {et~al.}(2015)\citenamefont
  {Fishman}, \citenamefont {Lee}, \citenamefont {Bord\'acs}, \citenamefont
  {K\'ezsm\'arki}, \citenamefont {Nagel},\ and\ \citenamefont
  {R{\~o}{\~o}m}}]{Fishman2015}%
  \BibitemOpen
  \bibfield  {author} {\bibinfo {author} {\bibfnamefont {R.~S.}\ \bibnamefont
  {Fishman}}, \bibinfo {author} {\bibfnamefont {J.~H.}\ \bibnamefont {Lee}},
  \bibinfo {author} {\bibfnamefont {S.}~\bibnamefont {Bord\'acs}}, \bibinfo
  {author} {\bibfnamefont {I.}~\bibnamefont {K\'ezsm\'arki}}, \bibinfo {author}
  {\bibfnamefont {U.}~\bibnamefont {Nagel}},\ and\ \bibinfo {author}
  {\bibfnamefont {T.}~\bibnamefont {R{\~o}{\~o}m}},\ }\bibfield  {title}
  {\bibinfo {title} {Spin-induced polarizations and nonreciprocal directional
  dichroism of the room-temperature multiferroic {${\mathrm{BiFeO}}_{3}$}},\
  }\href {https://doi.org/10.1103/PhysRevB.92.094422} {\bibfield  {journal}
  {\bibinfo  {journal} {Phys. Rev. B}\ }\textbf {\bibinfo {volume} {92}},\
  \bibinfo {pages} {094422} (\bibinfo {year} {2015})}\BibitemShut {NoStop}%
\bibitem [{\citenamefont {Katsura}\ \emph {et~al.}(2008)\citenamefont
  {Katsura}, \citenamefont {Onoda}, \citenamefont {Han},\ and\ \citenamefont
  {Nagaosa}}]{Katsura2008}%
  \BibitemOpen
  \bibfield  {author} {\bibinfo {author} {\bibfnamefont {H.}~\bibnamefont
  {Katsura}}, \bibinfo {author} {\bibfnamefont {S.}~\bibnamefont {Onoda}},
  \bibinfo {author} {\bibfnamefont {J.~H.}\ \bibnamefont {Han}},\ and\ \bibinfo
  {author} {\bibfnamefont {N.}~\bibnamefont {Nagaosa}},\ }\bibfield  {title}
  {\bibinfo {title} {Quantum theory of multiferroic helimagnets: Collinear and
  helical phases},\ }\href {https://doi.org/10.1103/PhysRevLett.101.187207}
  {\bibfield  {journal} {\bibinfo  {journal} {Physical Review Letters}\
  }\textbf {\bibinfo {volume} {101}},\ \bibinfo {eid} {187207} (\bibinfo {year}
  {2008})}\BibitemShut {NoStop}%
\end{thebibliography}
\end{document}